\documentclass[aps,prc,onecolumn,nopacs,preprintnumbers,nofootinbib,float]{revtex4-2}

\usepackage{epsfig}
\usepackage{color}
\usepackage[dvipsnames]{xcolor}
\usepackage{float,amsmath,amssymb}
\usepackage{graphicx}
\usepackage{subcaption}
\usepackage{url}
\usepackage{bbold}
\usepackage[utf8]{inputenc}
\usepackage{physics}
\usepackage{slashed}
\usepackage{tikz}

\newcommand{\xpom}{x_\mathbb{P}}

\newcommand{\gev}{\mathrm{GeV}}

\newcommand{\as}{\alpha_{\mathrm{s}}}
\newcommand{\xt}{{\mathbf{x}}}
\newcommand{\rt}{{\mathbf{r}}}

\newcommand{\kt}{{\mathbf{k}}}
\newcommand{\pt}{{\mathbf{p}}}

\newcommand{\bt}{{\mathbf{b}}}

\newcommand{\nc}{{N_\mathrm{c}}}
\newcommand{\nf}{{N_\mathrm{f}}}

\newcommand{\lo}{\mathrm{LO}}
\newcommand{\nlo}{\mathrm{NLO}}

\newcommand{\Ydip}{Y_\text{dip}}
\newcommand{\Yqqg}{Y_{\text{q}\bar{\text{q}}\text{g}}}
\newcommand{\Ybk}{{Y_{0,\text{BK}}}}
\newcommand{\zmin}{z_{\text{min}}}
\newcommand{\msbar}{$\overline{\text{MS}}$}

\newcommand{\acal}{\mathcal{A}}
\newcommand{\kcal}{\mathcal{K}}
\newcommand{\ocal}{\mathcal{O}}

\newcommand{\deltaD}{\delta_{(D)}}
\newcommand{\deltaDs}{\delta_{(D_s)}}
\newcommand{\epsilonDs}{\epsilon_{(D_s)}}

\usepackage[breaklinks,colorlinks,citecolor=citcolor,urlcolor=blue,linkcolor=lcolor]{hyperref}
\definecolor{lcolor}{rgb}{0.5,0,0}
\definecolor{citcolor}{rgb}{0,0.3,0.0}

\begin{document}

\title{Exclusive production of light vector mesons at next-to-leading order in the dipole picture}

\author{Heikki Mäntysaari}
\email{heikki.mantysaari@jyu.fi}
\author{Jani Penttala}
\email{jani.j.penttala@jyu.fi}
\affiliation{Department of Physics, University of Jyväskylä, P.O. Box 35, 40014 University of Jyv\"askyl\"a, Finland}
\affiliation{Helsinki Institute of Physics, P.O. Box 64, 00014 University of Helsinki, Finland}

\begin{abstract}
Exclusive production of light vector mesons in deep inelastic scattering is calculated at next-to-leading order in the dipole picture in the limit of high photon virtuality. The resulting expression is free of any divergences and suitable for numerical evaluations. The higher-order corrections are found to be numerically important, but they can be mostly captured by the nonperturbative fit parameters describing the initial condition for the small-$x$ evolution of the dipole scattering amplitude.
The vector meson production cross section is shown to depend only weakly  on the 
meson distribution amplitude and the factorization scale.  We also present phenomenological comparisons of our result to the existing exclusive $\phi$ and $\rho$ production data from HERA and find an excellent agreement at high virtualities.
\end{abstract}

\maketitle

\section{Introduction}

Deep inelastic scattering (DIS) is a powerful tool to study the partonic structure of protons and nuclei at high energies.
This process has been studied in detail in electron-proton collisions at HERA, where the vast amount of measured data has revealed a rapid increase in the density of gluons with small momentum fraction $x$~\cite{H1:2009pze, H1:2015ubc}. The observed increase cannot continue indefinitely without violating unitarity, and as such the saturation effects are expected to dominate the small-$x$ part of the hadron wave function. To describe QCD in this region of phase space where parton densities become of the same order as the inverse of the strong coupling, an effective field theory approach to QCD, called the Color Glass Condensate (CGC), has been developed~\cite{Iancu:2003xm,Gelis:2010nm,Blaizot:2016qgz}. 

In CGC, the high density of small-$x$ gluons gives rise to nonlinear dynamics that slows down the growth of the gluon density. 
Despite the success of the CGC-based calculations in describing various high-energy collider experiments~\cite{Morreale:2021pnn}, 
there has not been definitive experimental evidence of saturation. To get precise DIS data from the saturation region new experimental facilities have been proposed, such as the upcoming Electron-Ion Collider (EIC) in the US~\cite{Accardi:2012qut,Aschenauer:2017jsk,AbdulKhalek:2021gbh} and a similar collider at CERN~\cite{Agostini:2020fmq}. These facilities would allow for DIS measurements with heavy nuclei where the saturation effects are amplified approximately by $A^{1/3}$.
To meet the precision of these future experimental studies where nonlinear QCD dynamics is probed, it is necessary to promote the theory calculations in the CGC framework to higher-order accuracy.

One powerful process to probe gluon saturation is exclusive vector meson production as it requires an exchange of at least two gluons with the target. This renders the cross section roughly proportional to the gluon density squared~\cite{Ryskin:1992ui} at leading order (but the situation is more complicated at next-to-leading order in a collinear factorization based approach, see Ref.~\cite{Eskola:2022vpi}). Another advantage of it is that only in exclusive processes it is possible to measure the momentum transfer squared $t$ in the process. The momentum transfer dependence can be related to the impact parameter dependence via a Fourier transform, 
providing access to the spatial distribution of nuclear matter in nuclei at high energy~\cite{Klein:2019qfb,Mantysaari:2020axf} and to the generalized parton distribution functions~\cite{Belitsky:2005qn}.

A convenient approach for describing exclusive vector meson production in DIS is the dipole picture where the process can be written in terms of the virtual photon and meson light-front wave functions along with the dipole-target scattering amplitude~\cite{Kowalski:2006hc, Marquet:2007qa}. The dipole amplitude satisfies perturbative small-$x$ evolution equations, such as the Balitsky-Kovchegov (BK) equation~\cite{Balitsky:1995ub, Kovchegov:1999yj} which resums large logarithmic contributions $\sim \as \ln 1/x$. The photon wave function can be calculated perturbatively~\cite{Lepage:1980fj, Dosch:1996ss}, but the meson wave function is instead nonperturbative and therefore requires additional modeling. For heavy vector mesons one can take advantage of the small relative velocity of the quark-antiquark pair in the meson and model it as a fully nonrelativistic bound state~\cite{Ryskin:1992ui}, with velocity corrections that can be linked to the nonrelativistic QCD (NRQCD) matrix elements~\cite{Lappi:2020ufv}. Another possibility is to take the limit of high photon virtuality, $Q^2 \gg M_V^2$ (where $M_V$ is the meson mass), where one can make a twist expansion for the process~\cite{Chernyak:1983ej, Collins:1996fb}. 
This corresponds to writing the meson wave function in terms of a nonperturbative distribution amplitude, on top of which higher-order corrections can be calculated perturbatively.
This is especially suitable for light vector mesons and is a basic assumption in this paper.

The next-to-leading order (NLO) calculations in the dipole picture are starting to become available. First of all,  the BK equation is  available at NLO accuracy~\cite{Balitsky:2008zza, Lappi:2015fma, Lappi:2016fmu,Lappi:2020srm}. The NLO corrections to the virtual photon wave function have been calculated with both massless~\cite{Hanninen:2017ddy, Beuf:2016wdz, Beuf:2017bpd} and massive~\cite{Beuf:2021qqa, Beuf:2021srj} quarks. These developments enable phenomenological studies of proton and nuclear structure functions at small $x$, and also make it possible to determine the nonperturbative initial condition for the small-$x$ evolution of the dipole amplitude by performing fits to HERA data~\cite{Albacete:2010sy,Lappi:2013zma,Ducloue:2019jmy,Beuf:2020dxl}. Another recently proposed approach to determine the initial condition is based on a perturbative calculation of the proton color charge correlators in terms of the nonperturbative proton valence quark wave function~\cite{Dumitru:2020gla,Dumitru:2021tvw}.
In order to calculate exclusive vector meson production at NLO accuracy, the additional ingredient required is the meson wave function at NLO.
This wave function  has been calculated in the nonrelativistic limit for heavy vector mesons in Ref.~\cite{Escobedo:2019bxn} and used for calculating longitudinal heavy vector meson production at NLO accuracy in Ref.~\cite{Mantysaari:2021ryb} including also the first relativistic corrections~\cite{Lappi:2020ufv}.  Other recent developments towards the NLO accuracy in the CGC framework include, for example, studies of dijet production in DIS and hadronic collisions~\cite{Caucal:2021ent,Iancu:2020mos,Boussarie:2016ogo},  and inclusive hadron production in proton-lead collisions~\cite{Ducloue:2017dit,Ducloue:2017mpb,Ducloue:2016shw,Stasto:2013cha,Altinoluk:2014eka,Watanabe:2015tja,Iancu:2016vyg,Chirilli:2012jd}.

The main focus of this work, light vector meson production in the high-$Q^2$ limit at NLO, has been calculated in Ref.~\cite{Boussarie:2016bkq} using covariant perturbation theory in momentum space including nonlinear QCD dynamics in the shockwave approach.
In this paper, we calculate the NLO corrections using light-cone perturbation theory~\cite{Lepage:1980fj} in mixed transverse coordinate, longitudinal momentum fraction space.
The advantage of the light-cone perturbation theory is that the calculation can be divided into the photon and meson wave functions that need to be combined only at the end. One can also directly take advantage of the recently calculated photon NLO wave function. The mixed coordinate space is convenient as the transverse coordinates of the partons do not change during the interaction with the target at high energies.
Compared to Ref.~\cite{Boussarie:2016bkq} we also use a different scheme to subtract the rapidity divergence from the real gluon emission part. This  scheme  is developed in  Refs.~\cite{Ducloue:2017ftk,Iancu:2016vyg,Ducloue:2017mpb,Beuf:2014uia} in order to avoid unphysical results in single hadron production and in proton structure function calculations at NLO accuracy.
Our results are also straightforward to apply in phenomenological analyses using existing dipole amplitude fits as is demonstrated in this work.

  The paper is structured as follows. In Sec.~\ref{sec:exclusive_vm_production} we present the framework for vector meson production and explain the resummation of small-$x$ gluons. In Sec.~\ref{sec:wave_functions}, the photon and meson NLO wave functions are shown explicitly. The NLO corrections to the light vector meson wave function are calculated using light-cone perturbation theory at leading twist. We then proceed to calculate the production amplitude in Sec.~\ref{sec:meson_production_at_NLO} and present the result in the mixed space. In Sec.~\ref{sec:numerical_results}, we show numerical calculations of the NLO production amplitude along with comparisons to the existing $\rho$ and $\phi$ production data before presenting our conclusions in Sec.~\ref{sec:conclusions}.

\section{Exclusive scattering at high energy}
\label{sec:exclusive_vm_production}

\subsection{High energy factorization}

The scattering amplitude for exclusive vector meson production at high energy and in the zero squared momentum transfer $t=0$ limit can be written in a factorized form
\begin{multline}
    \label{eq:im_amplitude}
    -i \acal = \sum_f 2 \int \dd[D-2]{\xt_0} \dd[D-2]{\xt_1} \int \frac{\dd[]{z_0}\dd[]{z_1}}{(4\pi)^2} 4\pi\delta(z_0+z_1-1) \Psi^{\gamma^* \rightarrow q \bar q}_{f} \left(\Psi^{V \rightarrow q \bar q}_f\right)^* N_{01} \\
    +\sum_f 2\int \dd[D-2]{\xt_0} \dd[D-2]{\xt_1} \dd[D-2]{\xt_2} \int \frac{\dd[]{z_0}\dd[]{z_1}\dd[]{z_2}}{(4\pi)^3} 4\pi\delta(z_0+z_1+z_2-1) \Psi_{f}^{\gamma^* \rightarrow q \bar q g } \left(\Psi^{V \rightarrow q \bar q g}_f \right)^*  N_{012},
\end{multline}
and the coherent vector meson $V$ electroproduction cross section can now be obtained as 
\begin{equation}
\label{eq:dsigma_dt}
    \left.\frac{\dd\sigma^{\gamma^* + p \to V + p}}{\dd t}\right|_{t=0} = \frac{1}{16\pi} \left |\acal\right|^2.
\end{equation}
Here $\xt_{0,1,2}$ are the quark, antiquark and gluon transverse coordinates and $z_i$ denote the fractions of the photon's plus momentum carried by these partons.
This factorization is justified at high energy as the lifetimes of the virtual photon $q\bar q$ and $q\bar q g$ Fock states are much longer than the timescales related to the interactions with the target color field. We use the eikonal approximation and describe the interactions with the target in terms of Wilson line correlators. The Wilson line $V_{F,A}(\xt)$ describes a color rotation of a quark (fundamental representation $F$) or a gluon (adjoint representation $A$) when it propagates through the target, and the relevant correlators read
\begin{align}
    N_{01} &= 1-\frac{1}{\nc} \left\langle \Tr{V_F(\xt_0) V_F^\dagger(\xt_1)} \right\rangle \\
    \begin{split}
    N_{012} &= 1- \frac{1}{C_F \nc} \left\langle V_A^{ba}(\xt_2) \Tr{t^b V_F(\xt_0) t^a V_F^\dagger(\xt_1)}  \right\rangle \\
    & \approx 1 - \frac{\nc}{2C_F} \left( S_{02}S_{12} - \frac{1}{\nc^2} S_{01} \right),
    \end{split}
\end{align}
where $S_{01}=1-N_{01}$. Here we took the mean field limit where the average over the target color charge configurations denoted by $\langle \dotsm \rangle$ factorizes. These correlators 
satisfy the BK evolution equation discussed in Sec.~\ref{sec:bkevolution} and
depend implicitly on evolution rapidity which we will specify later.

We only consider forward production in this work even though this framework can also be extended to calculate the momentum transfer dependent cross section. 
The momentum transfer is the Fourier conjugate to the impact parameter, and thus being able to calculate the cross section at finite momentum transfer is an advantage of exclusive processes as this can be used to do spatial imaging of the hadron structure~\cite{Klein:2019qfb}. 
On the other hand, this means that calculating vector meson production at $t \neq 0$ would require us to implement a model to describe the nonperturbative spatial structure of the proton. As the purpose of this work is to focus on a rigorous NLO calculation of vector meson production cross section we choose not to employ any such modeling and limit our studies to $t=0$ where only the dipole amplitude integrated over the impact parameter is required. The same quantity is also probed in structure function measurements that are used to constrain the initial condition for the BK evolution of the dipole scattering amplitude $N_{01}$~\cite{Beuf:2020dxl}.

\subsection{Twist expansion}

The meson light-front wave function is highly nonperturbative. For heavy vector mesons one can model the wave function based on the nonrelativistic nature of heavy quarks~\cite{Lappi:2020ufv} but this simplification cannot be made for light mesons. On the other hand, the high-virtuality limit $Q^2 \gg M_V^2$, which is justified for light mesons, can be used to simplify the mathematical description of the process. In this limit transverse momentum scales on the meson side become corrections suppressed by powers of $1/Q^2$, leading to the twist expansion of the meson wave function~\cite{Chernyak:1983ej, Collins:1996fb}. The leading-twist term then does not depend on the transverse momentum scales of the meson, meaning that only the dependence on the longitudinal momenta remains.

The twist expansion can be explained formally using the virtual photon wave function. The photon wave function is exponentially suppressed in $Q^2 \rt^2$ where $\rt$ is the dipole size, which renders the relevant dipole sizes to be $\rt^2 \sim 1/Q^2$. We can then  do a Taylor expansion for the meson wave function $\Psi^V(\rt, z) = \Psi^V(0, z) + \frac{1}{6} \rt^2 \nabla_\rt^2 \Psi^V(0, z) + \ldots = \Psi^V(0, z) + \ocal\left(\frac{1}{Q^2}\right)$ (see also Ref.~\cite{Lappi:2020ufv}). Thus, only the dependence on the momentum fraction $z$ remains at leading order in the twist expansion. The momentum space equivalent of this is a delta function in terms of the quark transverse momentum $\kt$: $\Psi^V(\kt, z)  = (2\pi)^2 \delta^2(\kt) \Psi^V(\rt=0, z) + \ocal(\frac{1}{Q^2})$. This first term in the wave function corresponds to the twist-2 distribution amplitude $\phi(z)$ of the meson. In an NLO calculation the distribution amplitude has to be renormalized as we will demonstrate explicitly below, and the scale dependence of the renormalized distribution amplitude is described in terms of the Efremov-Radyushkin-Brodsky-Lepage (ERBL) evolution equation~\cite{Lepage:1980fj,Efremov:1979qk} which is discussed in more detail in Sec.~\ref{sec:erbl}.

The twist expansion also guarantees that we need the nonperturbative part of the meson wave function only for the $q \bar q$ state. The nonperturbative part for other Fock states, such as $q \bar q g$, is higher order in twist and can therefore be neglected at high $Q^2$ \cite{Ball:1998sk, Braun:1989iv} 
. This means that the meson wave function for the $q \bar q g$ state can be calculated perturbatively by considering a gluon emission from the $q \bar q $ state, i.e. at high virtualities the Fock state $ q\bar q g$ is created through the process $V \rightarrow q \bar q \rightarrow q \bar q g$ (see Figs.~\ref{fig:gluon_emission_quark} and \ref{fig:gluon_emission_antiquark}).

Another consequence of the high virtuality is that we need to consider only the longitudinal polarization for both the photon and the meson. A polarization flip is highly suppressed in coherent vector meson production such that the meson and photon effectively have the same polarization~\cite{H1:2009cml} (see also Ref.~\cite{Mantysaari:2020lhf}). In fact, in the limit of zero momentum exchange $t=0$ the polarization flip contribution vanishes exactly in our calculation. In the case of transverse production the leading-twist distribution amplitude is twist 3~\cite{Chernyak:1983ej}, meaning that transverse production is suppressed relative to longitudinal by $\sigma_T/\sigma_L \sim M_V^2 / Q^2$ for high virtualities.
Thus, total light vector meson production is given by the longitudinal cross section $\sigma(\gamma^*_L + A \rightarrow V_L + A)$ up to corrections of order $\ocal\left(\frac{M_V^2}{Q^2}\right)$.

\subsection{High-energy evolution}
\label{sec:bkevolution}

The dipole amplitude, given by the correlator $N_{01}=1-S_{01}$, satisfies the perturbative BK equation describing its energy dependence. At leading order the BK equation reads~\cite{Balitsky:1995ub, Kovchegov:1999yj}:
\begin{equation}
    \label{eq:BK}
    \frac{\partial }{\partial Y} S_{01} = \int \dd[2] \xt_2 K_\text{BK}(\xt_0,\xt_1,\xt_2) [S_{02} S_{12}-S_{01}].
\end{equation}
This equation is written in terms of a rapidity variable $Y$ which is discussed in more detail shortly. The kernel $K_\text{BK}$ describes the probability density for a dipole with transverse coordinates $\xt_0$ and $\xt_1$ to emit a gluon at the transverse coordinate $\xt_2$. Including the running coupling corrections following Ref.~\cite{Balitsky:2006wa}, the kernel can be written as
\begin{equation}
    \label{eq:bk-rc-balitsky}
  K_\text{BK}(\xt_0,\xt_1, \xt_2) = \frac{\nc \as(\xt_{01}^2)}{2\pi^2} \left[
        \frac{\xt_{01}^2}{\xt_{21}^2 \xt_{20}^2}  + \frac{1}{\xt_{20}^2} \left( \frac{\as(\xt_{20}^2)}{\as(\xt_{21}^2)} -1 \right) 
        + \frac{1}{\xt_{21}^2} \left( \frac{\as(\xt_{21}^2)}{\as(\xt_{20}^2)} -1 \right)
  \right]
\end{equation}
where we use the notation $\xt_{ij}=\xt_i - \xt_j$. The BK equation effectively resums the contributions $\as \ln 1/x \sim 1$ from small-$x$ gluons which is necessary for the stability of the perturbative calculations at high energy.

When higher-order corrections enhanced by large double transverse logarithms are resummed~\cite{Iancu:2015vea}, the NLO BK equation~\cite{Balitsky:2008zza} becomes stable and can in principle be used in phenomenological applications~\cite{Lappi:2016fmu}. 
A usual and numerically convenient approach, however, is to include resummations of the most important higher-order corrections to the leading-order BK equation. The leading-order BK equation with such resummations can be used to accurately approximate the full NLO BK equation~\cite{Lappi:2016fmu,Hanninen:2021byo}.
Several resummation schemes exist, and in this work the following equations are used (we adopt the terminology used in Ref.~\cite{Beuf:2020dxl}): \emph{KCBK}~\cite{Beuf:2014uia}, \emph{ResumBK}~\cite{Iancu:2015vea,Iancu:2015joa} and \emph{TBK}~\cite{Ducloue:2019ezk}. 
The nonperturbative initial conditions for these evolution equations have been determined in Ref.~\cite{Beuf:2020dxl} by performing a fit to the HERA structure function data~\cite{H1:2009pze}.
Of these, the evolution rapidity in the KCBK and ResumBK equations is the  projectile rapidity $Y=\ln \frac{k^+}{P^+}$, where $k^+$ and $P^+$ are the gluon and target plus momenta. 
We work in the frame where the photon plus momentum $q^+$ is large and the photon has no transverse momentum. The target plus momentum is obtained as $P^+ = Q_0^2/(2P^-)$, where the target transverse momentum scale is taken to be $Q_0^2=1\,\gev^2$ (note that the photon-nucleon center-of-mass energy reads $W^2=2q^+ P^-$).

Both KCBK (``kinematically constrained BK'') and ResumBK (``resummed BK'') involve a resummation of double transverse logarithms $\sim \as \ln \frac{|\xt_{02}|}{|\xt_{01}|}\ln \frac{|\xt_{12}|}{|\xt_{01}|}$, with ResumBK also resumming single transverse logarithms $\as \ln\frac{1}{\xt_{ij}^2 Q_s^2}$ at all orders. In the KCBK equation 
the double logarithms are resummed by explicitly requiring a time ordering between the subsequent gluon emissions, which results in a non-local equation. 
The third evolution equation, TBK (``BK equation in target rapidity''), uses the target rapidity $\eta$ as an evolution variable. This rapidity variable is related to the fraction of the target longitudinal momentum fraction transferred in the scattering process in the frame where the target has a large longitudinal momentum (see Ref.~\cite{Ducloue:2019ezk} for a detailed discussion). 
This evolution rapidity corresponds to 
\begin{equation}
    \eta = \ln \frac{1}{\xpom} = \ln\frac{W^2+Q^2}{Q^2+M_V^2}.
\end{equation}
Consequently, the TBK evolution can be thought as evolution in $\ln 1/\xpom$ whereas the KCBK and ResumBK equations correspond to evolution in $\ln W^2$.
In order to use dipole amplitudes as a function of the target rapidity $\eta$ in the impact factors written in terms of the projectile rapidity $Y$,  we use the same shift as in Ref.~\cite{Beuf:2020dxl}:
\begin{equation}
    \eta = Y - \ln \frac{1}{\min\{1, \xt_{01}^2 Q_0^2 \} }.
\end{equation}

The BK equation contains a transverse-coordinate dependent coupling constant. We model the running of the coupling in the coordinate space following Ref.~\cite{Beuf:2020dxl}:
\begin{equation}
    \label{eq:running_coupling}
    \as(\xt_{ij}^2)=\frac{4\pi}{\beta_0 \ln \left[ \left(\frac{\mu_0^2}{ \Lambda_\text{QCD}^2}\right)^{1/c}+\left(\frac{4C^2}{\xt_{ij}^2 \Lambda_\text{QCD}^2}\right)^{1/c} \right]^c}.
\end{equation}
This running coupling approaches a constant value in the infrared region $1/|\xt_{ij}| \gtrsim \Lambda_\text{QCD}$, with the constants $\mu_0$ and $c$ controlling its behavior there.  The values of these constants are chosen as in Ref.~\cite{Beuf:2020dxl}. The constant $C^2$ is a fit parameter that describes the relation between momentum and coordinate spaces, $\kt^2 = 4 C^2/\rt^2$, with the expected value $C^2 = e^{-2\gamma_E}$ from Fourier analysis~\cite{Kovchegov:2006vj,Lappi:2012vw}. The same coordinate space coupling constant is used when calculating the scattering amplitude, Eq.~\eqref{eq:im_amplitude}, where the coupling constant is included in the next-to-leading order photon and meson wave functions. As the  running coupling prescription~\eqref{eq:bk-rc-balitsky} can be seen to effectively choose the smallest of the three distance scales $\xt_{01}^2,\xt_{12}^2,\xt_{02}^2$, when calculating the $q\bar q g$ contribution in Eq.~\eqref{eq:im_amplitude} we choose to evaluate the coupling at the scale set by the smallest of the daughter dipoles,  as in  Ref.~\cite{Beuf:2020dxl}. When evaluating the $q\bar q$ term the scale choice is $\xt_{01}$.

\section{Light-front wave functions at next-to-leading order}
\label{sec:wave_functions}

The NLO corrections to exclusive vector meson production can be calculated in terms of the NLO wave functions for the photon and meson. In this section, we first list the relevant photon light-front wave functions at NLO accuracy calculated in Refs.~\cite{Hanninen:2017ddy, Beuf:2016wdz, Beuf:2017bpd}. Then, we proceed to calculate the light vector meson wave function at NLO in terms of the twist-2 distribution amplitude, and present the results Fourier transformed to mixed transverse coordinate, longitudinal momentum fraction space.

\subsection{On the regularization scheme}
\label{sec:reg_scheme}

The calculation will be done in two different regularization schemes. The first one is the conventional dimensional regularization (CDR) where the momenta and polarization vectors of all particles are continued to $D$ dimensions. The second one is the four-dimensional helicity (FDH) scheme where the polarization vectors are kept in four dimensions~\cite{Bern:1991aq, Bern:2002zk}. In our case, this amounts to real gluons having two polarization states.

To do the calculations simultaneously in both schemes we follow the notation of Ref.~\cite{Beuf:2021qqa}. The dimension arising from the gluon polarization vectors is denoted as $D_s$ to distinguish it from the dimension $D$ in the dimensional regularization. The CDR scheme corresponds to the case $D_s = D$, and for the FDH scheme we have $ D_s=4 $. Sums over gluon helicities can be calculated as $\sum_\lambda \epsilon_\lambda^{i} \epsilon_\lambda^{j*} = \deltaDs^{ij}$ where the subscript denotes that this Kronecker delta has $D_s-2$ transverse dimensions. In the sums over spin and Lorenz indices we take $D_s \geq D$ so that the following relations for the Kronecker deltas hold:
\begin{align}
    \deltaDs^{ij} \deltaDs^{ij} &= D_s-2  &
    \deltaD^{ij} \deltaD^{ij} &= D-2  &
    \deltaDs^{ij} \deltaD^{jk} &= \deltaD^{ik}.  
\end{align}
We will also make use of the following spinor identity~\cite{Hanninen:2017ddy}:
\begin{equation}
 \label{eq:spinor_identity}
    \bar u_{h'}(p-k) \slashed{\epsilon}^*_\lambda(k) u_h(p) 
    = \left( \bar v_{h'}(p-k) \slashed{\epsilon}_\lambda(k) v_h(p) \right)^*
    = \frac{2 p^+}{k^+} \sqrt{\frac{p^+}{p^+-k^+}}  \delta_{hh'} \epsilon_\lambda^{j*} \left[ \kt^i- \frac{k^+}{p^+} \pt^i \right]V_h^{ij}\left(\frac{k^+}{p^+}\right),
\end{equation}
where $h = \pm 1$ is the quark helicity and
\begin{equation}
     V^{ij}_h(z) = \left(1-\frac{z}{2}\right)\deltaDs^{ij} + i h \frac{z}{2} \epsilonDs^{ij}.
\end{equation}
Here the $D_s$-dimensional Levi-Civita tensor has to be understood through the Fierz identity
\begin{equation}
     \epsilonDs^{ij} \epsilonDs^{kl} = \deltaDs^{ik} \deltaDs^{jl} - \deltaDs^{jk} \deltaDs^{il}.
\end{equation}
This identity is valid in $D_s$ dimensions if there are no more than two Levi-Civita tensors~\cite{Hanninen:2017ddy}, which holds for the calculations considered in this paper.
  
\subsection{Photon wave function}
\label{sec:photon_wf}

The photon light-front wave functions in the massless quark case have been calculated in Refs.~\cite{Hanninen:2017ddy, Beuf:2016wdz, Beuf:2017bpd} and are shown here for completeness. In our notation, an additional factor of $\frac{1}{2q^+}\prod_i \frac{1}{\sqrt{z_i}}$ appears in the wave functions. This additional factor follows from a different choice of the integration measure which we choose to be $\prod_i \frac{\dd[2]{\xt_{i}}\dd{z_i}}{4\pi}$. With this choice the leading-order wave function for the virtual photon in the mixed transverse coordinate and plus momentum fraction space is 
\begin{equation}
\label{eq:photon_LO}
    \Psi^{\gamma^* \rightarrow q \bar q}_{f,\lo}(z_0,\xt_{01}) = \frac{e e_f Q}{\pi} \delta_{\alpha_0 \alpha_1} \delta_{h_0,-h_1} z_0(1-z_0) K_{\frac{D-4}{2}}\left( |\xt_{01}| \overline Q\right)
    \times \left( \frac{\overline Q}{2\pi |\xt_{01}| } \right)^{\frac{D-4}{2}}.
\end{equation}
We always work in the frame where the photon transverse momentum is zero.
Here $e_f$ is the fractional charge of the quark with flavor $f$, $Q^2$ is the virtuality of the photon, $z_i=k_i^+/q^+$ is the (anti)quark's fraction of the photon plus momentum,  and $\alpha_i$ and $h_i$ are the color and helicity indices. We also use the short-hand notation $\overline Q^2 = z_0(1-z_0)Q^2$. Quantities corresponding to the quark are denoted with $i=0$ and antiquark with $i=1$. We note that the  last factor, which is equal to 1 at $D=4$, is absent in Ref.~\cite{Hanninen:2017ddy} where the transverse momenta of the observed particles are kept in two dimensions. Here ``observed" particles are those which appear as the final state in the wave function, not including soft or collinear particles. In this paper, we choose to evaluate the transverse momenta of the observed particles in $D-2$ dimensions,
as this is necessary for regularizing the NLO meson wave function. However, this term does not have any contribution to the final cross section where all $\frac{1}{D-4}$  divergences have been canceled. In principle, this factor multiplied by $K^{\gamma_L^*}$ contributes a finite logarithm term $\sim \ln |\xt_{01}|$. It however cancels when we perform the UV subtraction in Sec.~\ref{sec:UV_subtraction}.

The next-to-leading order correction to the photon wave function can be written as
\begin{equation}
\label{eq:photon_qq_NLO}
    \Psi^{\gamma^* \rightarrow q \bar q}_{f,\nlo}(z_0,\xt_{01}) = \frac{e e_f Q}{\pi} \delta_{\alpha_0 \alpha_1} \delta_{h_0,-h_1} z_0(1-z_0)  K_{\frac{D-4}{2}}\left( |\xt_{01}| \overline Q\right) \frac{\as C_F}{ 2\pi} K^{\gamma_L^*}
    \times \left( \frac{\overline Q}{2\pi |\xt_{01}|} \right)^{\frac{D-4}{2}}
\end{equation}
where
\begin{equation}
    K^{\gamma_L^*} = \left[ \frac{3}{2} + \ln(\frac{\alpha^2}{z_0(1-z_0)}) \right] \left[ \frac{2}{4-D} +\gamma_E +\ln(\pi \xt_{01}^2 \mu^2 ) \right] + \frac{1}{2} \ln^2 \left(\frac{z_0}{1-z_0}\right) -\frac{\pi^2}{6} + \frac{5}{2}+ \frac{1}{2}\frac{D_s-4}{D-4}.
\end{equation}
Here $\alpha$ is the infrared cut-off for the gluon plus momentum fraction and $\mu$ is the mass scale for dimensional regularization, and the last term depends on the regularization scheme.

The virtual photon wave function for the Fock state $q \bar q g$ can be written as
\begin{equation}
\label{eq:photon_qqg}
    \Psi^{\gamma^* \rightarrow q \bar q g}_f(z_i,\xt_i) = 4 e e_f Q g t^a_{\alpha_0 \alpha_1}  \delta_{h_0,-h_1} \frac{1}{\sqrt{z_2}} \epsilon_{h_2}^{j*} \left[ z_1(1-z_1) V^{ij}_{h_0}\left( \frac{z_2}{z_0+z_2} \right) I^i_{(l)} - z_0(1-z_0) V^{ij}_{-h_0}\left( \frac{z_2}{z_1+z_2} \right) I^i_{(m)} \right]
\end{equation}
where
\begin{equation}
\begin{aligned}
    I^i_{(l)} &= I^i\left(\xt_{102}, \xt_{20}, \overline Q^2_{(l)}, \omega_{(l)} \right) & I^i_{(m)} &= I^i\left(\xt_{012}, \xt_{21}, \overline Q^2_{(m)}, \omega_{(m)}\right) \\
    \overline Q^2_{(l)} &= z_1(1-z_1) Q^2  &  \overline Q^2_{(m)} &= z_0(1-z_0) Q^2 = \overline Q^2 \\
    \omega_{(l)} &= \frac{z_0 z_2}{z_1(z_0+z_2)^2} & \omega_{(m)} &= \frac{z_1 z_2}{z_0 (z_1+z_2)^2} \\
    \xt_{ijk} &= \xt_{ij}- \frac{z_k}{z_j+z_k} \xt_{kj},
\end{aligned}
\end{equation}
and
\begin{equation}
    I^i \left(\bt, \rt, \overline Q^2, \omega \right) = (4 \pi^2 \mu \rt^2)^{\frac{4-D}{2}} \frac{i}{8\pi^2} \frac{\rt^i}{\rt^2}
     \int_0^\infty \dd{u} u^{1-D/2} e^{-u \overline Q^2} e^{-\frac{\bt^2}{4u}} \Gamma\left( 1+\frac{D-4}{2}, \frac{\omega \rt^2}{4u} \right).
\end{equation}
Quantities with the subscript 2 correspond to the emitted gluon.
The function $I^i$ differs from the similar one in Ref.~\cite{Hanninen:2017ddy} by an additional power $\frac{4-D}{2}$ for the variable $u$, which has the same origin as the last factor in Eqs.~\eqref{eq:photon_LO} and~\eqref{eq:photon_qq_NLO}.

\subsection{Light vector meson wave function}
\label{sec:meson_wf}

In this section, we  calculate the NLO corrections to the light vector meson light-front wave function. The calculation is  done in the limit where the transverse coordinate dependence of the meson leading-order wave function can be neglected. As discussed in Sec.~\ref{sec:exclusive_vm_production}, this follows from the large photon photon virtuality. %
This means that we can neglect all mass scales in the meson, allowing us to set the meson mass $M_V$ to zero along with the transverse momenta $\kt_i$ of the quark and the antiquark. We work in a frame where both the photon and the vector meson transverse momenta are zero, as we consider forward production. Consequently, at leading order the meson wave function is given by a delta function in the transverse plane:
\begin{equation}
\label{eq:meson_qq_LO_mom}
   \Psi^{V \rightarrow q \bar q}_{f,\lo}(z_0, \kt_0) = c_f \frac{\delta_{\alpha_0 \alpha_1}}{\sqrt{N_c}} \delta_{h_0, -h_1} \frac{\pi f_V}{e_V\sqrt{N_c}} \phi_0(z_0)  \times (2 \pi)^{D-2} \delta^{D-2}(\kt_0) .
\end{equation}
Here $\phi_0(z)$ is the (bare) distribution amplitude of the meson that describes how the meson plus momentum is shared by the two quarks. This wave function is normalized in such a way that it gives the correct decay constant $f_V$ given that the distribution amplitude is normalized as 
\begin{equation}
\label{eq:normalization}
    \int_0^1 \dd{z} \phi_0(z) = 1.
\end{equation}
The decay constant $f_V$ is related to the leptonic width by    
\begin{equation}
\label{eq:leptonic_width}
    \Gamma(V \rightarrow l^+ l^-) = \frac{4\pi \alpha_\mathrm{em}^2 f_V^2}{3 M_V}.
\end{equation} 

The wave function \eqref{eq:meson_qq_LO_mom} describes the probability of the meson to split into a quark-antiquark pair with the flavor $f$. 
Here $c_f$ is a normalization factor needed for mesons that consist of a superposition of different flavored quark-antiquark states. 
For example, the $\rho$ meson can be written at leading order as $\ket{\rho}= \frac{1}{\sqrt{2}} \left( \ket{u \bar u} - \ket{d \bar d} \right)$, giving us $c_u = \frac{1}{\sqrt{2}}$ and $c_d = -\frac{1}{\sqrt{2}}$. The normalization factors are also related to the effective charge fraction of the meson which is defined by $e_V = \sum_f c_f e_f$. We emphasize that in the high-$Q^2$ limit the dependence on the vector meson type is included in the nonperturbative distribution amplitude $\phi_0(z_0)$ (in addition to the normalization factors $f_V, e_V$ and $c_f$).

\begin{figure}
	\centering
    \begin{subfigure}{0.45\textwidth}
        \centering
        \includegraphics[width=\textwidth]{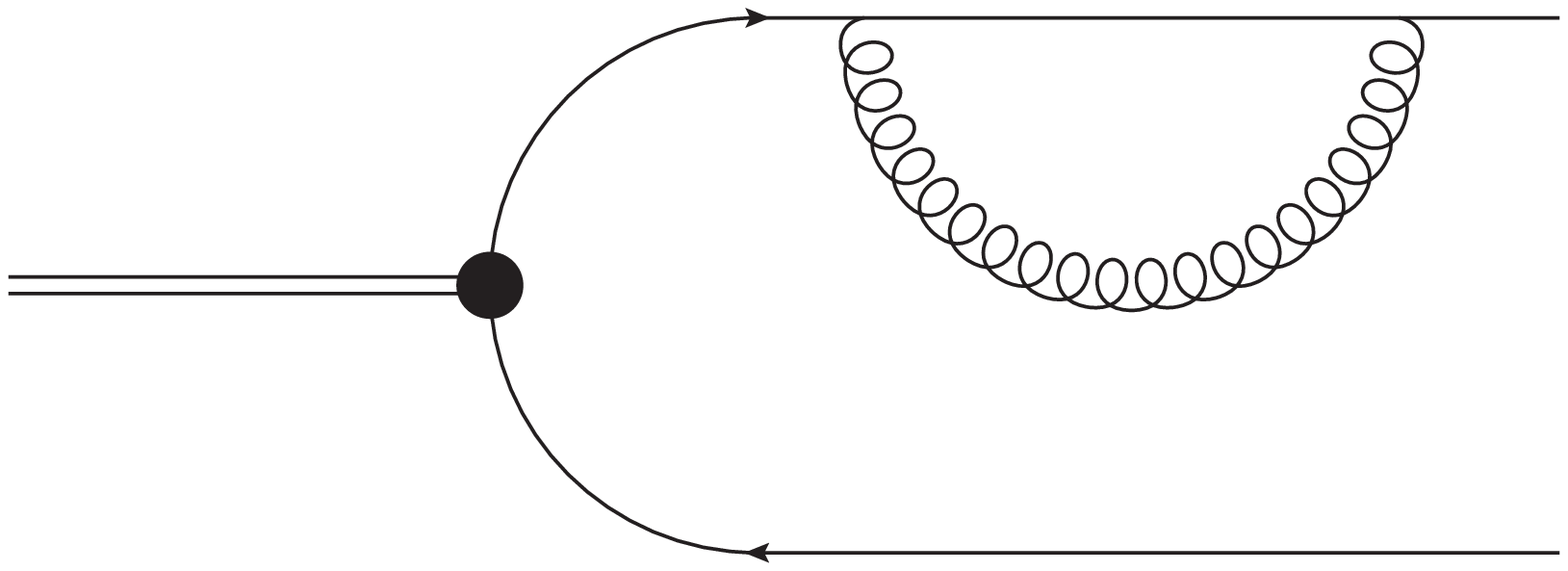}
        \begin{tikzpicture}[overlay]
         \node[anchor=south east] at (4.2cm,2.65cm) {$\kt_0, z_0$};
         \node[anchor=south east] at (4.2cm,0.5cm) {$\kt_1, z_1$};
         \end{tikzpicture}
        \caption{ }
        \label{fig:quark_self-energy}
    \end{subfigure}
    \begin{subfigure}{0.45\textwidth}
        \centering
        \includegraphics[width=\textwidth]{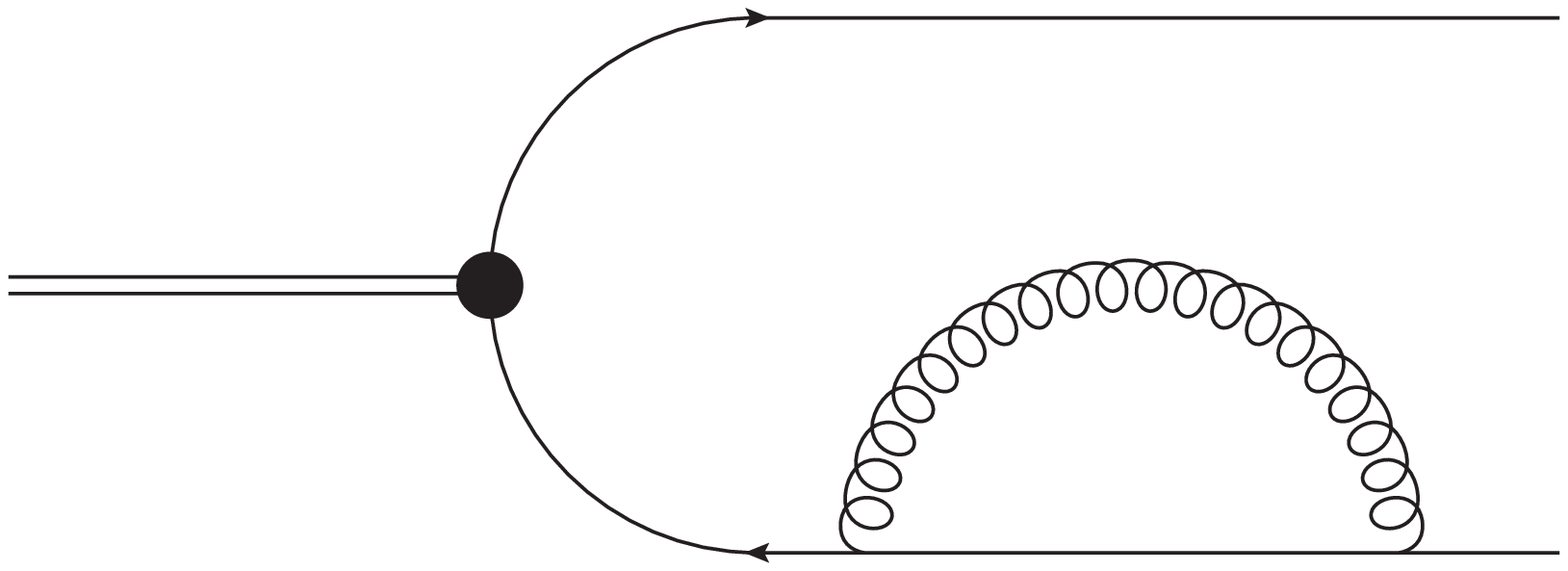}
        \begin{tikzpicture}[overlay]
         \node[anchor=south east] at (4.2cm,2.65cm) {$\kt_0, z_0$};
         \node[anchor=south east] at (4.2cm,0.5cm) {$\kt_1, z_1$};
         \end{tikzpicture}
        \caption{ }
        \label{fig:antiquark_self-energy}
    \end{subfigure}
    \begin{subfigure}{0.45\textwidth}
        \centering
        \includegraphics[width=\textwidth]{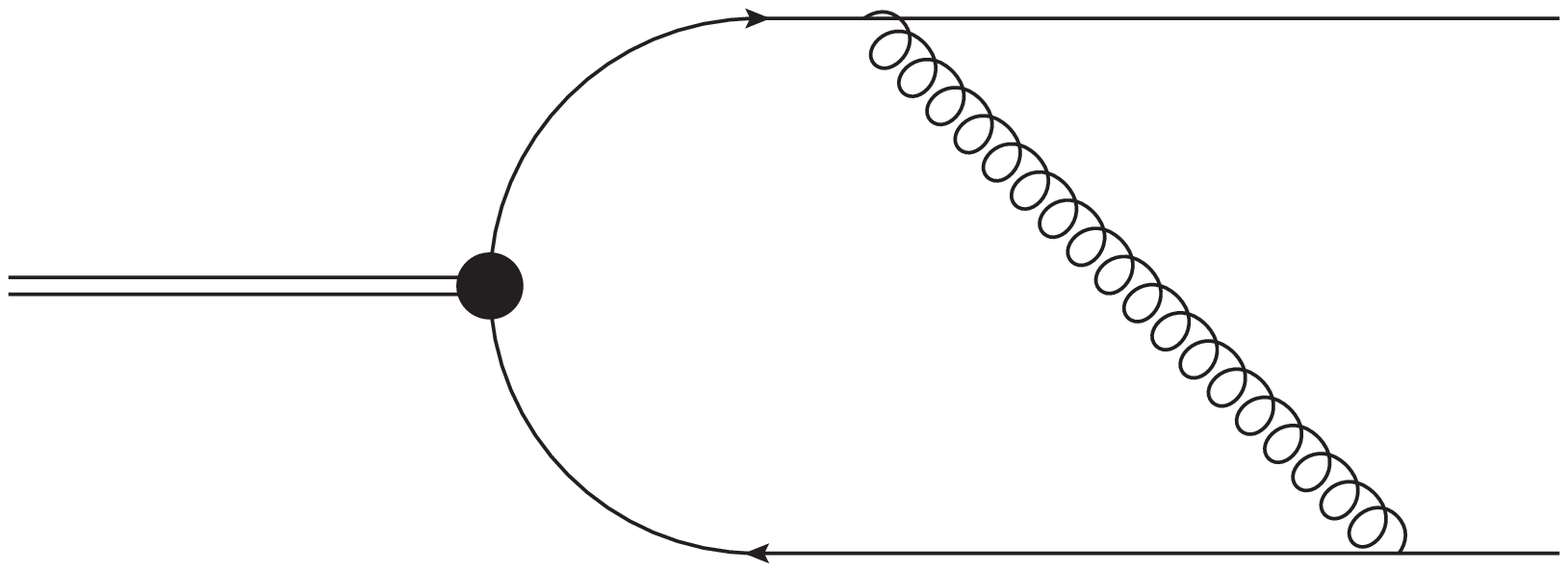}
        \begin{tikzpicture}[overlay]
         \node[anchor=south east] at (4.2cm,2.6cm) {$\kt_0, z_0$};
         \node[anchor=south east] at (4.2cm,0.5cm) {$\kt_1, z_1$};
         \node[anchor=south east] at (0.4cm,2.6cm) {$\kt_0', z_0'$};
         \node[anchor=south east] at (0.4cm,0.5cm) {$\kt_1', z_1'$};
         \node[anchor=south east] at (2.8cm,1.9cm) {$\kt_2, z_2$};
         \end{tikzpicture}
        \caption{ }
        \label{fig:vertex_correction_1}
    \end{subfigure}
    \begin{subfigure}{0.45\textwidth}
        \centering
        \includegraphics[width=\textwidth]{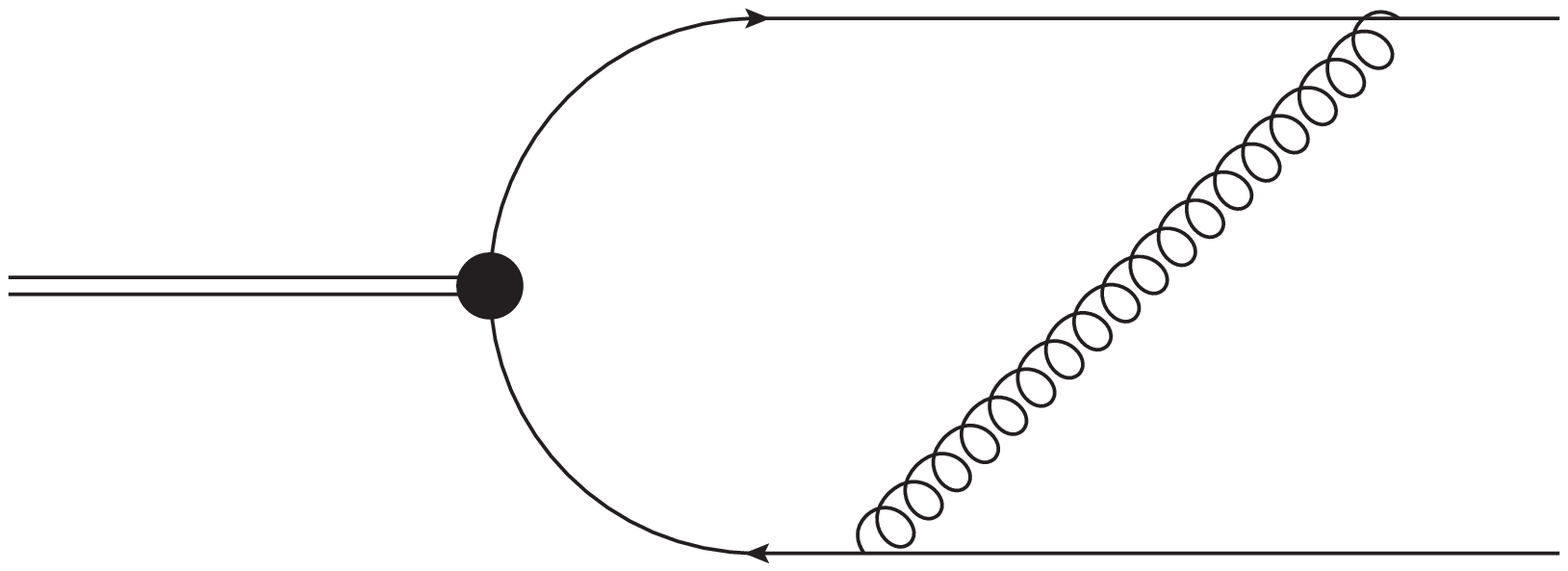}
        \begin{tikzpicture}[overlay]
         \node[anchor=south east] at (4.2cm,2.6cm) {$\kt_0, z_0$};
         \node[anchor=south east] at (4.2cm,0.5cm) {$\kt_1, z_1$};
         \node[anchor=south east] at (0.4cm,2.6cm) {$\kt_0', z_0'$};
         \node[anchor=south east] at (0.4cm,0.5cm) {$\kt_1', z_1'$};
         \node[anchor=south east] at (1.7cm,1.9cm) {$\kt_2, z_2$};
         \end{tikzpicture}
        \caption{ }
        \label{fig:vertex_correction_2}
    \end{subfigure}
    \begin{subfigure}{0.45\textwidth}
        \centering
        \includegraphics[width=\textwidth]{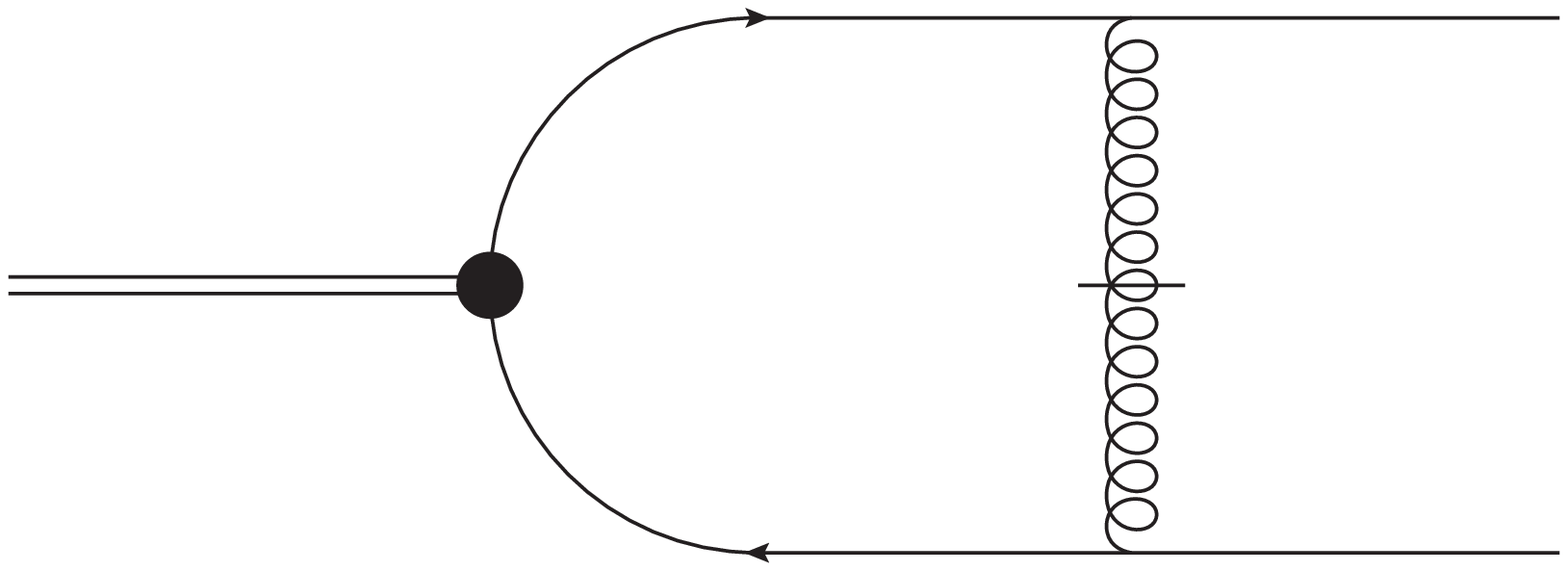}
        \begin{tikzpicture}[overlay]
         \node[anchor=south east] at (4.2cm,2.6cm) {$\kt_0, z_0$};
         \node[anchor=south east] at (4.2cm,0.5cm) {$\kt_1, z_1$};
         \node[anchor=south east] at (0.4cm,2.6cm) {$\kt_0', z_0'$};
         \node[anchor=south east] at (0.4cm,0.5cm) {$\kt_1', z_1'$};
         \end{tikzpicture}
        \caption{ }
        \label{fig:vertex_correction_virtual}
    \end{subfigure}
    
    \begin{subfigure}{0.45\textwidth}
        \centering
        \includegraphics[width=\textwidth]{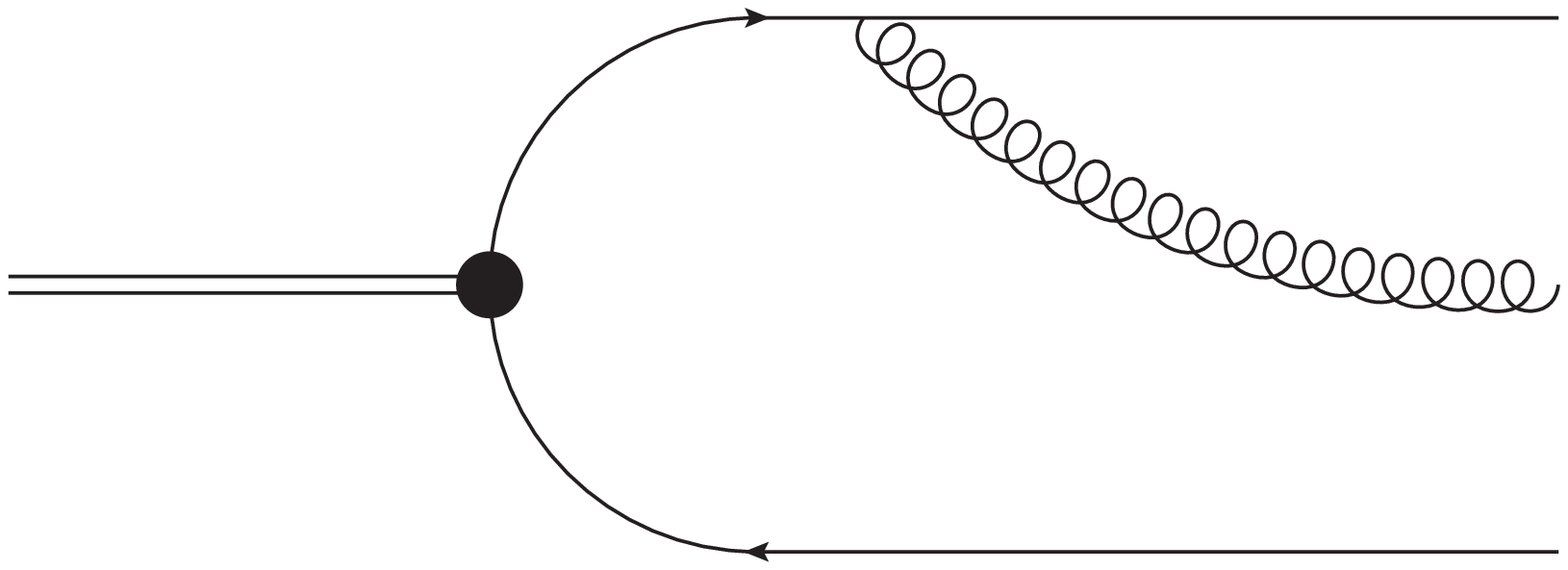}
        \begin{tikzpicture}[overlay]
         \node[anchor=south east] at (4.2cm,2.6cm) {$\kt_0, z_0$};
         \node[anchor=south east] at (4.2cm,0.5cm) {$\kt_1, z_1$};
         \node[anchor=south east] at (4.2cm,2.0cm) {$\kt_2, z_2$};
         \node[anchor=south east] at (0.4cm,2.6cm) {$\kt_0', z_0'$};
         \end{tikzpicture}
        \caption{ }
        \label{fig:gluon_emission_quark}
    \end{subfigure}
    \begin{subfigure}{0.45\textwidth}
        \centering
        \includegraphics[width=\textwidth]{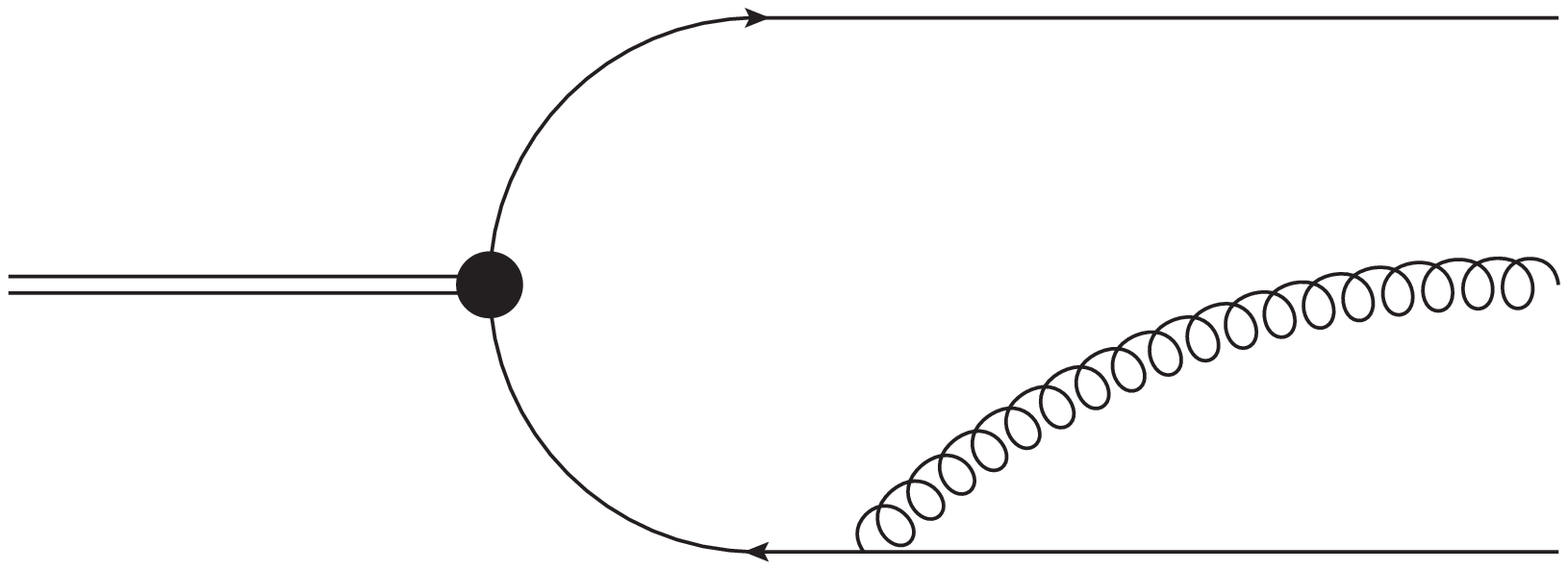}
        \begin{tikzpicture}[overlay]
         \node[anchor=south east] at (4.2cm,2.6cm) {$\kt_0, z_0$};
         \node[anchor=south east] at (4.2cm,0.5cm) {$\kt_1, z_1$};
         \node[anchor=south east] at (4.2cm,2.0cm) {$\kt_2, z_2$};
         \node[anchor=south east] at (0.4cm,0.5cm) {$\kt_1', z_1'$};
         \end{tikzpicture}
        \caption{ }
        \label{fig:gluon_emission_antiquark}
    \end{subfigure}
	  \caption{NLO corrections to the meson light-front wave function}
        \label{fig:Feynman_diagrams}
\end{figure}

At next-to-leading order, we get  perturbative corrections to the meson wave function from Feynman diagrams shown  in Fig.~\ref{fig:Feynman_diagrams}. Of these, the diagrams~\ref{fig:quark_self-energy} and \ref{fig:antiquark_self-energy}, corresponding to the self-energy corrections of the quark and antiquark, evaluate to zero. This is a consequence of the dimensional regularization used in the calculation, as these diagrams give transverse integrals with no mass scales (in the high-$Q^2$ limit considered here  where we neglect the quark and meson masses and the quark transverse momenta) such as $\int \frac{\dd[D-2]{\kt_0}}{(2\pi)^{D-2}} \frac{1}{\kt^2_0} = 0$.

To calculate the rest of the diagrams we use the Feynman rules of the light-cone perturbation theory from Ref.~\cite{Hanninen:2017ddy}.  For the diagram~\ref{fig:vertex_correction_1} this gives:
\begin{equation}
   \label{eq:vertex_correction_1} 
   \begin{split}
   \Psi_f^{\ref{fig:vertex_correction_1}}=&
   \int \frac{\dd[D-2]{\kt_0'} \dd{k_0^{\prime +}}}{(2\pi)^{D-2}4\pi}
   \frac{1}{4 k_2^+ k_0^{\prime +} k_1^{\prime +}\left(P^- - k_0^{-}-k_1^{-} \right)\left(P^- - k_0^{-}-k_1^{\prime -} \right)} \Psi_\lo^{V \rightarrow q \bar q}(z_0', \kt_0'; \alpha_{0}', \alpha_{1}', h_{0}', h_{1}') \sqrt{\frac{z'(1-z')}{z(1-z)}} \\
   &\times \mu^{4-D} g^2 t_{\alpha_0 \alpha_{0}'}^{\alpha_2} t_{\alpha_{1}' \alpha_1}^{\alpha_2}  \bar u(0) \slashed{\epsilon}^*_{h_2}(2) u(0') \bar v(1') \slashed{\epsilon}_{h_2}(2) v(1) \\
   =&-4 \pi \frac{\as C_F}{2\pi} \frac{\mu^{4-D}}{\kt_0^2} c_f \frac{\delta_{\alpha_0 \alpha_1}}{\sqrt{N_c}} \delta_{h_0, -h_1} \frac{\pi f_V}{e_V\sqrt{N_c}}  \\
   &\times \int^1_{z_0+\alpha}\dd{z'} \phi_0(z') \frac{z_0}{z'} \frac{1}{(z'-z_0)^2} \left[  z' (1-z') +z_0(1-z_0) +\frac{D_s-4}{2} (z'-z_0)^2 \right],
   \end{split}
\end{equation}
where the identity~\eqref{eq:spinor_identity} has been used to simplify the result. The square root factor in the first line comes from our choice for the integration measure, and the quark and antiquark transverse momenta after the gluon exchange are $\kt_0$ and $\kt_1=-\kt_0$. We use a notation $u(0) = u_{h_0}(k_0),v(1) = v_{h_1}(k_1)$ for the quark and antiquark spinors, and the primed quantities 
correspond to the intermediate quark and antiquark whose spins and helicities are summed over
(see Fig.~\ref{fig:Feynman_diagrams}).

The contribution of the diagram~\ref{fig:vertex_correction_2} is similar to the diagram~\ref{fig:vertex_correction_1}. An explicit calculation gives the result Eq.~\eqref{eq:vertex_correction_1} with the substitutions $z_0 \rightarrow 1-z_0$ and $z' \rightarrow 1-z'$:
\begin{equation}
   \label{eq:vertex_correction_2} 
\begin{split}
   \Psi_f^{\ref{fig:vertex_correction_2}}=& -4 \pi \frac{\as C_F}{2\pi} \frac{\mu^{4-D}}{\kt_0^2} c_f \frac{\delta_{\alpha_0 \alpha_1}}{\sqrt{N_c}} \delta_{h_0, -h_1} \frac{\pi f_V}{e_V\sqrt{N_c}}\\
   &\times\int^{z_0-\alpha}_{0} \dd{z'} \phi_0(z') \frac{1-z_0}{1-z'} \frac{1}{(z'-z_0)^2} \left[  z' (1-z') +z_0(1-z_0) +\frac{D_s-4}{2} (z'-z_0)^2 \right].
\end{split}
\end{equation}
Here we used the symmetry condition $\phi(z')=\phi(1-z')$ which follows from the parity of the vector meson.

The final contribution to the NLO $q \bar q$ wave function comes from the diagram~\ref{fig:vertex_correction_virtual} describing an exchange of an instantaneous gluon between the quark and the antiquark. The contribution from this diagram can be evaluated to give
\begin{equation}
    \label{eq:vertex_correction_virtual}
    \Psi_f^{\ref{fig:vertex_correction_virtual}}= 8 \pi \frac{\as C_F}{2\pi} \frac{\mu^{4-D}}{\kt_0^2}c_f \frac{\delta_{\alpha_0 \alpha_1}}{\sqrt{N_c}} \delta_{h_0, -h_1} \frac{\pi f_V}{e_V\sqrt{N_c}} \int^{1}_{0} \dd{z'} \phi_0(z')  \frac{z_0(1-z_0)}{(z'-z)^2} 
    \Big[  \theta\left(z'-z_0-\alpha \right) + \theta\left(z_0-z'-\alpha \right) \Big].
\end{equation}
Summing these contributions together, we get the NLO correction to the meson $q \bar q$ wave function:
\begin{equation}
\begin{split}
\label{eq:meson_qq_NLO_mom}
    \Psi_{f,\nlo}^{V \rightarrow q \bar q}(z_0, \kt_0) =& c_f \frac{\delta_{\alpha_0 \alpha_1} }{\sqrt{N_c}} \delta_{h_0, -h_1}
    \frac{\pi f_V}{e_V\sqrt{N_c}}\frac{ \as C_F}{2\pi} \\ 
    &\times4\pi \frac{\mu^{4-D}}{\kt_0^2} \int^{1}_0 \dd{z'} \phi_0(z') \left[ \theta(z'-z_0-\alpha) \frac{z_0}{z'} \left(1+\frac{1}{z'-z_0} \right)+\theta(z_0-z'-\alpha) \frac{1-z_0}{1-z'} \left(1+\frac{1}{z_0-z'} \right) \right.\\
    &\left.+\frac{D_s-4}{2} \left( \frac{1-z_0}{1-z'} \theta(z_0-z') +\frac{z_0}{z'}\theta(z'-z_0)\right)\right].
\end{split}
\end{equation}
It should be noted that this NLO correction does not affect the normalization~\eqref{eq:normalization} of the distribution amplitude. The reason for this is that the decay constant is given by $f_V \sim  \int \dd{z_0} \int \dd[D-2]{\kt_0} \Psi^{V \rightarrow q \bar q}(z_0, \kt_0)$, and this integral vanishes for Eq.~\eqref{eq:meson_qq_NLO_mom} in dimensional regularization.

We also need the wave function for the $q \bar q g$ state. This is simply given by the sum of the  diagrams~\ref{fig:gluon_emission_quark} and \ref{fig:gluon_emission_antiquark}, which evaluates to: 
\begin{equation}
\begin{split}
    \label{eq:meson_qqg_mom}
    \Psi_f^{V \rightarrow q \bar q g}(z_i, \kt_i) =& c_f \frac{\pi f_V}{e_V\sqrt{N_c}}  \frac{2gt^a_{\alpha_0 \alpha_1}}{\sqrt{N_c z_2}} 
     \epsilon^{j*}_{h_2} \delta_{h_0,-h_1}\mu^{\frac{4-D}{2}}\frac{\kt_2^i}{\kt_2^2} \\
    & \times \left[(2\pi)^{D-2} \delta^{D-2}(\kt_1) \phi_0(z_1) V_{h_0}^{ij}\left( \frac{z_2}{z_0+z_2}\right)
    -(2\pi)^{D-2} \delta^{D-2}(\kt_0) \phi_0(z_0) V_{-h_0}^{ij}\left( \frac{z_2}{z_1+z_2}\right) \right].
\end{split}
\end{equation}
Note that the momentum conservation implies $z_0+z_1+z_2=1$ and $\kt_0+\kt_1+\kt_2=0$.

These wave functions are presented in the momentum space. For the meson production calculation we need the mixed space wave functions which can be calculated from the momentum space wave functions by a Fourier transform in the transverse plane. The leading-order wave function in the mixed space is given by
\begin{equation}
\label{eq:meson_LO}
\begin{split}
    \Psi_{f,\lo}^{V \rightarrow q \bar q}(z_0, \xt_{01}) =& \int \frac{\dd[D-2]{\kt_0}\dd[D-2]{\kt_1}}{(2\pi)^{2(D-2)}} e^{i (\kt_0 \vdot \xt_{0} + \kt_1 \vdot \xt_1)} (2\pi)^{D-2} \delta^{D-2}(\kt_0+\kt_1) \Psi_\lo^{V \rightarrow q \bar q}(z_0, \kt_0) \\
    &= 
    c_f\frac{\delta_{\alpha_0 \alpha_1}}{\sqrt{N_c}} \delta_{h_0,-h_1}
    \frac{\pi f_V}{e_V\sqrt{N_c}}\phi_0(z_0).
\end{split}
\end{equation}
The NLO correction to the $q \bar q$ wave function is given by 
\begin{equation} 
\begin{split}
\label{eq:meson_qq_NLO}
    \Psi_{f,\nlo}^{V \rightarrow q \bar q}(z_0, \xt_{01}) =& \int \frac{\dd[D-2]{\kt_0}\dd[D-2]{\kt_1}}{(2\pi)^{2(D-2)}} e^{i (\kt_0 \vdot \xt_{0} + \kt_1 \vdot \xt_1)} (2\pi)^{D-2} \delta^{D-2}(\kt_0+\kt_1) \Psi_\nlo^{V \rightarrow q \bar q}(z_0, \kt_0) \\
    =&c_f \frac{\delta_{\alpha_0 \alpha_1}}{\sqrt{N_c}} \delta_{h_0,-h_1}
    \frac{\pi f_V}{e_V\sqrt{N_c}}\frac{ \as C_F}{2\pi} \left( \pi \xt_{01}^2 \mu^2 \right)^{\frac{4-D}{2}} \Gamma\left(\frac{D-4}{2}\right) \\
    &\times \int^{1}_0 \dd{z'} \phi_0(z') \left[ \theta(z'-z_0-\alpha) \frac{z_0}{z'} \left(1+\frac{1}{z'-z_0} \right)+\theta(z_0-z'-\alpha) \frac{1-z_0}{1-z'} \left(1+\frac{1}{z_0-z'} \right)  \right.\\
    &\left.+\frac{D_s-4}{2} \left( \frac{1-z_0}{1-z'} \theta(z_0-z') +\frac{z_0}{z'}\theta(z'-z_0)\right)\right],
\end{split}
\end{equation}
and the wave function for the $q \bar qg$ state is
\begin{equation}
\begin{split}
\label{eq:meson_qqg}
    &\Psi_f^{V \rightarrow q \bar q g}(z_i, \xt_i) =\int \frac{\dd[D-2]{\kt_0}\dd[D-2]{\kt_1}\dd[D-2]{\kt_2}}{(2\pi)^{3(D-2)}}e^{i( \kt_0 \vdot \xt_{0} +\kt_1 \vdot \xt_{1} +\kt_2 \vdot \xt_{2})  } (2\pi)^{D-2} \delta^{D-2}(\kt_0+\kt_1+\kt_2) \Psi^{V \rightarrow q \bar q g}(z_i, \kt_i)\\
    &= c_f \frac{\pi f_V}{e_V\sqrt{N_c}} \frac{2g t^a_{\alpha_0 \alpha_1}}{\sqrt{N_c z_2}}  \epsilon^{j*}_{h_2} \delta_{h_0, -h_1} \mu^{\frac{4-D}{2}}\int \frac{\dd[D-2]{\kt_2}}{(2\pi)^{D-2}}  \frac{\kt_2^i}{\kt_2^2} \left[ \phi_0(z_1) V_{h_0}^{ij}\left( \frac{z_2}{z_1+z_2} \right) e^{i \kt_2 \vdot \xt_{20}}- \phi_0(z_0) V_{-h_0}^{ij}\left( \frac{z_2}{z_0+z_2} \right) e^{i \kt_2 \vdot \xt_{21}} \right] \\
    &= c_f \frac{\pi f_V}{e_V\sqrt{N_c}}\frac{2g t^a_{\alpha_0 \alpha_1}}{\sqrt{N_c z_2}}  \epsilon^{j*}_{h_2} \delta_{h_0, -h_1}  \left[ \phi_0(z_1) V_{h_0}^{ij}\left( \frac{z_2}{z_0+z_2} \right) J^i(\xt_{20})
    - \phi_0(z_0) V_{-h_0}^{ij}\left( \frac{z_2}{z_1+z_2} \right) J^i(\xt_{21}) \right],
\end{split}
\end{equation}
where
\begin{equation}
    J^i(\rt) = \frac{i}{2\pi} \frac{\rt^i}{\rt^2} \left( \pi \mu \rt^2 \right)^{\frac{4-D}{2}} \Gamma\left(1+ \frac{D-4}{2} \right).
\end{equation}

\section{Light vector meson production at next-to-leading order}
\label{sec:meson_production_at_NLO}

\subsection{Production amplitude}

Having determined the NLO corrections to the meson wave function, we now have all the ingredients to calculate the exclusive light meson production amplitude. We substitute the photon wave functions for the $q \bar q$ (sum of Eqs.~\eqref{eq:photon_LO} and \eqref{eq:photon_qq_NLO}) and $q \bar qg$ (Eq.~\eqref{eq:photon_qqg}) states, along with the meson wave functions for the $q \bar q$ (sum of Eqs.~\eqref{eq:meson_LO} and \eqref{eq:meson_qq_NLO}) and $q \bar qg$ (Eq.~\eqref{eq:meson_qqg}) states, into Eq.~\eqref{eq:im_amplitude} to obtain the production amplitude and keep terms up to $\mathcal{O}(\as)$. The production amplitude can then be divided into the dipole ($q\bar q$) and real emission ($q \bar q g$) parts. The dipole part contains the leading-order result 
\begin{equation}
\label{eq:amp_LO}
    -i \acal_\lo = 
    \frac{e  Q f_V}{\pi}  \int_0^1 \dd{z_0} \int \dd[D-2]{\xt_{01}} \int \dd[D-2]{\bt}  N_{01}  z_0 (1-z_0) K_{\frac{D-4}{2}}\left( |\xt_{01}|\overline Q\right) \phi_0(z_0)\times \left( \frac{\overline Q}{2\pi |\xt_{01}|} \right)^{\frac{D-4}{2}},
\end{equation}
and the NLO correction
\begin{equation}
\label{eq:amp_qq_NLO}
    \begin{split}
    &-i \acal^{q \bar q}_\nlo = \frac{e Q f_V}{\pi}   \frac{\as C_F}{2\pi}\int_0^1\dd{z_0}\int \dd[D-2]{\xt_{01}} \int \dd[D-2]{\bt}N_{01}  z_0 (1-z_0) K_{\frac{D-4}{2}}\left( \xt_{01}\overline Q\right) \times \left( \frac{\overline Q}{2\pi |\xt_{01}|} \right)^{\frac{D-4}{2}}  \\
    & \times \left\{ \left( \pi \xt_{01}^2 \mu^2 \right)^{\frac{4-D}{2}} \Gamma\left(\frac{D-4}{2}\right) \int^{1}_0 \dd{z'} \phi_0(z') \left[ \theta(z'-z_0-\alpha) \frac{z_0}{z'} \left(1+\frac{1}{z'-z_0} \right)+\theta(z_0-z'-\alpha) \frac{1-z_0}{1-z'} \left(1+\frac{1}{z_0-z'} \right)  \right]\right.\\
    &\left. + \phi_0(z_0)  K^{\gamma_L^*} +\frac{D_s-4}{D-4} \int_0^1 \dd{z'} \phi_0(z') \left[ \frac{1-z_0}{1-z'} \theta(z_0-z') +\frac{z_0}{z'}\theta(z'-z_0)\right]\right\},
    \end{split}
\end{equation}
where $\bt = (\xt_0 + \xt_1)/2$.
The real emission part reads
\begin{equation}
\begin{split}
\label{eq:amp_qqg}
    -i \acal^{q \bar qg} =& \frac{e  Q f_V}{\pi}  \frac{\as C_F}{2\pi} \int_0^1\dd{z_0} \int \dd[D-2]{\xt_{01}} \int \dd[D-2]{\bt} \int_\alpha^{1-z_0} \dd{z_2} \int \dd[D-2]{\xt_{20}}  N_{012}  \\
    &\times \frac{-8\pi^2}{z_2} \left\{
    \phi_0(z_1) J^i(\xt_{20}) \frac{1}{1-z_1} \left[
        z_1 \left( z_0^2+(1-z_1)^2 \right) I_{(l)}^i - z_0 \left( z_0(1-z_0)+z_1(1-z_1) \right) I_{(m)}^i
    \right] \right. \\
    &\left.+\phi_0(z_0) J^i(\xt_{21}) \frac{1}{1-z_0} \left[
        z_0 \left( z_1^2+(1-z_0)^2 \right) I_{(m)}^i - z_1 \left( z_0(1-z_0)+z_1(1-z_1) \right) I_{(l)}^i
    \right] \right. \\
    &+\left. \frac{D_s-4}{2} z_2^2 \left[ \phi_0(z_1) \frac{z_1}{1-z_1} J^i(\xt_{20}) I^i_{(l)} + \phi_0(z_0) \frac{z_0}{1-z_0} J^i(\xt_{21}) I^i_{(m)}   \right]
    \right\}
\end{split}
\end{equation}
where  $\bt=z_0 \xt_0 + z_1 \xt_1 + z_2 \xt_2$. These choices for the impact parameter $\bt$ follow Ref.~\cite{Beuf:2020dxl}, but we note that in the $t=0$ case the weighting of the coordinates by the momentum fractions $z_i$ in the definition of $\bt$ does not affect the results.

The coherent vector meson $V$ electroproduction cross section \eqref{eq:dsigma_dt} can now be evaluated using the scattering amplitude
\begin{equation}
    i\acal = i \acal^{q \bar q}_\lo + i \acal^{q \bar q}_\nlo +  i \acal^{q \bar qg}.
\end{equation}
When squaring the amplitude, we keep terms up to $\mathcal{O}(\as)$. However, the $i\acal^{q\bar qg}$ amplitude also contains a large contribution enhanced by a large logarithm $\ln 1/z_2 \sim 1/\as$, and as such this contribution has to be considered as being part of the leading order amplitude. This is in practice done by taking into account the BK evolution as we will discuss in more detail in Secs.~\ref{sec:softgluon} and~\ref{sec:numerical_results}.

\subsection{UV subtraction}
\label{sec:UV_subtraction}
The dipole ($-i \acal^{q \bar q}_\nlo$) and real emission ($-i \acal^{q \bar qg}$) parts of the amplitude are separately UV divergent. However, most of the divergences cancel in their sum. Therefore it is useful to subtract the UV divergent part of the real emission and combine it with the dipole part. The subtracted term is chosen to be

\begin{equation}
\begin{split}
\label{eq:UV_sub}
    -i \acal^{q \bar qg}_\text{UV} =& \frac{e Q f_V}{\pi}  \frac{\as C_F}{2\pi} \int_0^1\dd{z_0} \int \dd[D-2]{\xt_{01}} \int \dd[D-2]{\bt} \int_\alpha^{1-z_0} \dd{z_2} \int \dd[D-2]{\xt_{20}}  N_{01}  \\
    &\times\frac{-8\pi^2}{z_2} \left\{
    \phi_0(z_1) J^i(\xt_{20})I_\text{UV}^i\left(\xt_{20}, z_1(1-z_1)Q^2\right) \frac{z_1}{1-z_1}
        \left( z_0^2+(1-z_1)^2 \right) \right.\\
    &\left.+\phi_0(z_0) J^i(\xt_{21})I_\text{UV}^i\left(\xt_{21},z_0(1-z_0) Q^2\right) \frac{z_0}{1-z_0}\left( z_1^2+(1-z_0)^2 \right) 
    \right. \\
    &+\left. \frac{D_s-4}{2} z_2^2 \left[ \phi_0(z_1) \frac{z_1}{1-z_1} J^i(\xt_{20}) I^i_{(l)} + \phi_0(z_0) \frac{z_0}{1-z_0} J^i(\xt_{21}) I^i_{(m)}   \right]
    \right\}
\end{split}
\end{equation}
where
\begin{equation}
    I_\textrm{UV}^i\left(\rt, \overline Q^2\right) = \frac{i}{4\pi^2} (\pi\mu \rt^2)^{\frac{4-D}{2}} \frac{\rt^i}{\rt^2} \Gamma\left(1+\frac{D-4}{2}\right) e^{-\frac{\rt^2}{\xt_{01}^2 e^{\gamma_E}}}  K_{\frac{D-4}{2}}\left(|\xt_{01}| 
    \overline Q \right)\times \left( \frac{\overline Q}{2\pi |\xt_{01}|} \right)^{\frac{D-4}{2}}.
\end{equation}
This choice for the UV subtraction term is analogous to the one in Ref.~\cite{Hanninen:2017ddy} and also what is used when considering heavy vector meson production in Ref.~\cite{Mantysaari:2021ryb}. Unlike in Ref.~\cite{Hanninen:2017ddy}, we choose to include the additional factor $ \left( \frac{\overline Q}{2\pi |\xt_{01}|} \right)^{\frac{D-4}{2}}$ to the UV subtraction to cancel the same factor in the dipole part.

The integrals over $\xt_{20}$ and $z_2$ can be done analytically, which simplifies the UV subtraction term to
\begin{equation}
\begin{split}
\label{eq:UV_sub2}
    -i \acal^{q \bar qg}_\text{UV} =& - \frac{e Q f_V}{\pi}  \frac{\as C_F}{2\pi} \int_0^1\dd{z_0} \int \dd[D-2]{\xt_{01}} \int \dd[D-2]{\bt} N_{01} \phi_0(z_0) z_0(1-z_0)  K_{\frac{D-4}{2}}\left( \xt_{01} \overline Q\right)\times \left( \frac{\overline Q}{2\pi |\xt_{01}|} \right)^{\frac{D-4}{2}} \\
    & \times \left\{\Gamma\left(1+\frac{D-4}{2}\right) \Gamma\left(\frac{4-D}{2}\right) (\pi\mu^2 \xt_{01}^2 e^{\gamma_E})^{\frac{4-D}{2}} \left[ 3 +2 \ln(\frac{\alpha^2}{z_0(1-z_0)}) \right] +\frac{D_s-4}{D-4} \right\}.
\end{split}
\end{equation}
We then add this to the dipole part, which gives us
\begin{equation}
    \begin{split}
    \label{eq:NLO_qq_UV_subtracted}
    -i \acal^{q \bar q}_\textrm{sub} =& \frac{e Q f_V}{\pi}  \int_0^1\dd{z_0}\int \dd[D-2]{\xt_{01}} \int \dd[D-2]{\bt}N_{01}  z_0 (1-z_0)  K_{\frac{D-4}{2}}\left( \xt_{01} \overline Q\right)\times \left( \frac{\overline Q}{2\pi |\xt_{01}|} \right)^{\frac{D-4}{2}} \\
    &\times \int_0^1 \dd{z'} \phi_0(z') \Bigg\{ 
        \delta(z_0-z') \\
        &\left.+\frac{\as C_F}{2\pi} \left[
            K(z_0,z')  \left( \frac{2}{D-4}-\ln(\pi \mu^2 \xt_{01}^2 e^{\gamma_E}) \right)+ \delta(z_0-z') \left( \frac{1}{2} \ln^2\left(\frac{z_0}{1-z_0}\right)-\frac{\pi^2}{6}+\frac{5}{2}\right)
        \right]\right. \\
        &\left. +\frac{\as C_F}{2\pi}\frac{D_s-4}{D-4} \left[-\frac{1}{2} \delta(z_0-z')+\frac{1-z_0}{1-z'} \theta(z_0-z') +\frac{z_0}{z'}\theta(z'-z_0) \right]
    \right\}.
    \end{split}
\end{equation}
Here $K(z, z')$ is the kernel of the ERBL equation~\cite{Lepage:1980fj,Efremov:1979qk}
which describes the scale dependence of the distribution amplitude as we will discuss in Sec.~\ref{sec:erbl}:
\begin{equation}
    \label{eq:ERBL}
    K(z,z') 
    =\frac{z}{z'} \left( 1+\frac{1}{z'-z} \right) \theta(z'-z-\alpha)+ \frac{1-z}{1-z'} \left( 1+ \frac{1}{z-z'} \right) \theta(z-z'-\alpha) + \left(\frac{3}{2} +\ln(\frac{\alpha^2}{z(1-z)}) \right) \delta(z'-z).
\end{equation}
This form for the ERBL kernel is equivalent to the usual one written in terms of the plus distributions in the limit $\alpha \rightarrow 0$.

After the UV subtraction, the real emission part becomes finite and reads:
\begin{equation}
\begin{split}
\label{eq:NLO_qqg_UV_subtracted}
    -i \acal^{q \bar qg}_\text{sub} =& \frac{e  Q f_V}{\pi}  \frac{\as C_F}{2\pi} \int_0^1\dd{z_0} \int \dd[2]{\xt_{01}} \int \dd[2]{\bt} \int_\alpha^{1-z_0} \dd{z_2} \int \dd[2]{\xt_{20}}   \\
    &\times\frac{2}{\pi z_2} \phi_0(z_0)\left\{ N_{012} K_0(RQ) \frac{1}{1-z_0} \left[
        z_0 \left( (1-z_0-z_2)^2+(1-z_0)^2 \right)\frac{1}{\xt_{21}^2} 
        \right.\right. \\
        &\left. \left.- (1-z_0-z_2) \left( z_0(1-z_0)+ (1-z_0-z_2)(z_0+z_2) \right) \frac{\xt_{20} \vdot \xt_{21}}{\xt_{20}^2 \xt_{21}^2}
    \right]
    \right. \\
    &\left.-N_{01} \frac{z_0}{1-z_0} \left( (1-z_0-z_2)^2+(1-z_0)^2 \right) \frac{1}{\xt_{21}^2} e^{-\frac{\xt_{21}^2}{\xt_{01}^2 e^{\gamma_E}}}  K_0\left( \xt_{01} \overline Q\right)   \right\},
\end{split}
\end{equation}
where $R^2 = z_0 z_1 \xt_{01}^2+z_1 z_2 \xt_{21}^2 +z_0 z_2 \xt_{20}^2$. 

\subsection{ERBL evolution and the renormalized distribution amplitude}
\label{sec:erbl}
The dipole part $ -i \acal^{q \bar q}_\text{sub}$, Eq.~\eqref{eq:NLO_qq_UV_subtracted}, still contains a divergence of the form $\frac{1}{D-4}$ which is canceled when the distribution amplitude is renormalized. We define the renormalized distribution amplitude $\phi(z, \mu_F)$ as
\begin{equation}
\label{eq:DA_renormalization}
    \phi(z,\mu_F) = \phi_0(z) + \frac{\as C_F}{2\pi} \int_0^1 \dd[]{z'} K(z,z') \phi_0(z') \left( \frac{2}{D-4} + \gamma_E -\ln(4\pi) +\ln(\frac{\mu_F^2}{\mu^2}) \right),
\end{equation}
where $\mu_F$ is the factorization scale. This choice for the finite terms in the subtraction corresponds to the \msbar \ scheme. 
We note that the distribution amplitude depends on the regularization scheme (FDH or CDR), as in practice it has to be determined from some experimental process for which an NLO calculation also depends on the same scheme choice. In principle the scheme-dependent term $\sim (D_s-4)/(D-4)$ in Eq.~\eqref{eq:NLO_qq_UV_subtracted} could be also included in the definition of the renormalized distribution amplitude \eqref{eq:DA_renormalization}. However, in this work we choose to keep the scheme dependence explicitly visible in the dipole term, Eq.~\eqref{eq:NLO_qq_UV_subtracted}. This allows us to  straightforwardly quantify the scheme dependence which is shown in Appendix~\ref{app:scheme_dependence} to be negligible.

The renormalized distribution amplitude satisfies the ERBL evolution equation~\cite{Lepage:1980fj,Efremov:1979qk}
\begin{equation}
    \frac{\partial \phi(z, \mu_F)}{\partial \ln \mu_F^2} = \frac{\as C_F}{2 \pi}  \int_0^1 \dd{z'} K(z,z') \phi(z', \mu_F),
\end{equation}
where the kernel $K(z,z')$ is given in Eq.~\eqref{eq:ERBL}.
We note that this renormalization does not change Eq.~\eqref{eq:normalization} for the normalization of the distribution amplitude as the $z$-integral over the ERBL kernel vanishes: $\int_0^1 \dd[]{z} K(z,z')=0$.

Next we use Eq.~\eqref{eq:DA_renormalization} to write the bare distribution amplitude $\phi_0(z)$ in $-i \acal^{q \bar q}_\textrm{sub}$, Eq.~\eqref{eq:NLO_qq_UV_subtracted}, in terms of the renormalized distribution amplitude. We also choose to use the scale dependent renormalized distribution amplitude instead of the bare distribution in the NLO part, as their difference is now formally higher order in $\as$. This results in the finite expression
\begin{equation}
    \begin{split}
    \label{eq:NLO_qq_fin}
    &-i \acal^{q \bar q}_\textrm{fin} = \frac{e Q f_V}{\pi}   \int_0^1\dd{z_0} \int \dd[2]{\xt_{01}} \int \dd[2]{\bt}N_{01}  z_0 (1-z_0)  K_0\left( \xt_{01} \overline Q\right) \\
    &\times \int_0^1 \dd{z'} \phi(z', \mu_F) \left\{ 
        \delta(z_0-z') +\frac{\as C_F}{2\pi} \left[
            -K(z_0,z')  \ln( \frac{\mu_F^2\xt_{01}^2 e^{2\gamma_E}}{4} )+\delta(z_0-z') \left( \frac{1}{2} \ln^2\left(\frac{z_0}{1-z_0}\right)-\frac{\pi^2}{6}+\frac{5}{2}\right)
        \right]\right. \\
    &\left. +\frac{\as C_F}{2\pi}\frac{D_s-4}{D-4} \left[-\frac{1}{2} \delta(z_0-z')+\frac{1-z_0}{1-z'} \theta(z_0-z') +\frac{z_0}{z'}\theta(z'-z_0) \right]
    \right\}.
    \end{split}
\end{equation}
Similarly we can replace $\phi_0(z)$ by $\phi(z,\mu_F)$ in the real emission part \eqref{eq:NLO_qqg_UV_subtracted}.

Let us briefly consider the evolution of the renormalized distribution amplitude. It is useful to write the distribution amplitude in terms of the eigenfunctions $f_n(z)$ of the ERBL kernel 
\begin{equation}
    \int_0^1 \dd[]{z'} K(z,z') f_n(z') = \lambda_n f_n(z).
\end{equation}
The eigenfunctions can be written in terms of the Gegenbauer polynomials $C^{(\frac{3}{2})}_n$ as $f_n(z)= 6z(1-z) C^{(\frac{3}{2})}_n(2z-1)$, and the corresponding eigenvalues are given by~\cite{Lepage:1980fj}
\begin{equation}
\label{eq:lambda_n}
    \lambda_n=-\frac{1}{2}+\frac{1}{(n+1)(n+2)}-2\sum_{k=2}^{n+1} \frac{1}{k}. 
\end{equation}
Writing the distribution amplitude as a sum of the eigenfunctions, the ERBL equation then tells us that the coefficients in the sum depend on the factorization scale:
\begin{equation}
\label{eq:gegenbauer_expansion}
    \phi(z, \mu_F) =  \sum_{n=0}^\infty a_n(\mu_F) f_n(z).
\end{equation}
Taking into account the running of the coupling constant as $\as(\mu_F^2) = \frac{4\pi}{\beta \ln(\mu_F^2/\Lambda_\textrm{QCD}^2)}$, we can solve the evolution of the coefficients $a_n$ explicitly~\cite{Lepage:1980fj}:
\begin{equation}
\label{eq:an_evolution}
    a_n(\mu_F) = a_n \ln(\frac{\mu_F^2}{\Lambda_\textrm{QCD}^2})^{\frac{2 C_F}{\beta} \lambda_n}.
\end{equation}
Here $\Lambda_\textrm{QCD}=0.241\,\gev$ and $\beta=(11\nc - 2\nf)/3$ with  $\nf=3$.
Values for the coefficients $a_n$ at an initial scale are a nonperturbative input for the calculation. These coefficients also depend on the considered vector meson, and should be determined from experimental data.
It should be noted that the eigenvalue $\lambda_n$ is zero for the term $n=0$ and negative for the $n>0$ terms. This means that the first term is actually constant in $\mu_F$, and the higher-order terms become suppressed as $\mu_F$ increases. In the asymptotic limit $\mu_F \rightarrow \infty$ only the first term contributes, and the distribution amplitude then simplifies to $\phi(z, \mu_F=\infty) = 6z(1-z)$. Here we have also used the fact that the coefficient $a_0$ of the first term is actually determined by the normalization condition Eq.~\eqref{eq:normalization}, as the orthogonality of the Gegenbauer polynomials guarantees that only the first term contributes to the normalization, giving us $a_0 =1$.
It should also be noted that parity conservation demands that the distribution amplitude is invariant under the substitution $z \leftrightarrow 1-z$, meaning that all terms with $n = \mathrm{odd}$ are zero in the sum.

We point out that Eq.~\eqref{eq:an_evolution} is divergent for $\mu_F = \Lambda_\textrm{QCD}$. In practice, we avoid this singularity by introducing an infrared (IR) cutoff $\mu_{F0}$ for the ERBL evolution and freeze the distribution amplitude below this scale: $\phi(z,\mu_F) = \phi(z, \mu_{F0})$ for $\mu_F < \mu_{F0}$. We choose the value of the IR cut-off to be $\mu_{F0}= 1 \, \gev$. The dependence on the IR cutoff is quantified in Appendix~\ref{app:scheme_dependence}.

\subsection{Soft gluon divergence}
\label{sec:softgluon}
The real emission part still has an IR divergence from the lower limit $\alpha$ of the $z_2$ integral. This is related to the emission of soft gluons from the dipole, and to the rapidity evolution of the dipole amplitude. This can be seen by noting that the singular part of the real emission can be written as
\begin{equation}
\begin{split}
\label{eq:qqg_singularity}
    -i \acal^{q \bar qg}_\text{sing}
    =& \frac{e  Q f_V}{\pi}  \int_0^1\dd{z_0} \int \dd[2]{\xt_{01}} \int \dd[2]{\bt} \phi(z_0,  \mu_F) z_0(1-z_0)  K_0\left( |\xt_{01}| \overline Q\right)   \\
    &\times\frac{\as C_F}{2\pi}\int_{\zmin}^{1-z_0} \dd{z_2} \int \dd[2]{\xt_{20}}\frac{2}{\pi z_2} \left[ N_{012}-N_{01} \right]   \frac{\xt_{01}^2}{\xt_{20}^2 \xt_{21}^2}
\end{split}
\end{equation}
where the identity~\cite{Hanninen:2017ddy}
\begin{equation}
    \int \dd[2]{\xt_{2}} \left[ \frac{\xt_{01}^2}{\xt_{20}^2 \xt_{21}^2} -\frac{1}{\xt_{20}^2} e^{-\xt_{20}^2/(\xt_{01}^2 e^{\gamma_E})}-\frac{1}{\xt_{21}^2} e^{-\xt_{21}^2/(\xt_{01}^2 e^{\gamma_E})} \right] = 0
\end{equation}
has been used.
Note that as we do not have an explicit dependence on the infrared cutoff $\alpha$ in the integrands anymore, from now on the lower limit of the $z_2$ integral is denoted by $\zmin$ whose value will be discussed shortly.
We can recognize the integrand in Eq.~\eqref{eq:qqg_singularity} as the kernel of the (fixed coupling leading order) BK equation~\eqref{eq:BK}. This can then be combined with the leading-order term ($\as^0$ part of Eq.~\eqref{eq:NLO_qq_fin}), and the sum of these two contributions corresponds to using in the leading-order term a dipole amplitude evolved from the initial rapidity $Y_0$ to the  rapidity
\begin{equation}
    \Ydip = Y_0 + \ln\frac{1-z_0}{\zmin}.
\end{equation}

At finite center-of-mass energy the lower limit 
$\zmin$ of the $z_2$ integral should not be taken to zero. In particular, the invariant mass of the $q\bar q g$ system should be much less than $W^2$ in order to justify the usage of the eikonal approximation, which imposes the lower limit $\zmin$.
We follow Refs.~\cite{Mantysaari:2021ryb,Beuf:2020dxl} and choose
\begin{equation}
\label{eq:zmin}
 \zmin = \min \left( e^{Y_0} \frac{Q_0^2}{W^2+Q^2-m_N^2}, 1-z_0 \right).   
\end{equation}
Here the minimum comes from the kinematic constraint $z_0 + z_2 \leq 1$ which guarantees that the dipole does not evolve backwards in rapidity. As we are interested in the high (but finite) energy limit, the minimum is only needed in a small subset of the integration region and in practice the evolved rapidity is
\begin{equation}
    \Ydip = \ln((1-z_0)\frac{W^2+Q^2-m_N^2}{Q_0^2}).
\end{equation}

For the $\as$-suppressed terms the dependence on the evolution rapidity is formally of higher order in the coupling constant. Following again Refs.~\cite{Mantysaari:2021ryb,Beuf:2020dxl} we choose to use the same evolution rapidity $\Ydip$ when evaluating the next-to-leading order terms in the dipole part, Eq.~\eqref{eq:NLO_qq_fin}. The $z_2$-dependent evolution rapidity used with real gluon emission term is obtained from the definition $Y= \ln \frac{k^+}{P^+}$ and can be written as~\cite{Mantysaari:2021ryb}
\begin{equation}
    \Yqqg = \ln z_2 + \ln \frac{W^2 + Q^2 - m_N^2}{Q_0^2}.
\end{equation}

\subsection{Full result}
We can now write the scattering amplitude for light meson electroproduction in its full form. It reads
\begin{equation}
	\label{eq:total_NLO}
		-i\acal =  \frac{e Q f_V}{\pi} \int \dd[2]{\xt_{01}}\int \dd[2]{\bt} \int_0^1 \dd{z_0}
		  \Bigg\{ \kcal_{q \bar q}^\lo(Y_0) 
		+\frac{\alpha_s C_F}{2\pi}  \kcal_{q \bar q}^\nlo(\Ydip) 
		+ \int \dd[2]{\xt_{20}} \int_{\zmin}^{1-z_0} \dd[]{z_2} \frac{\alpha_s C_F}{2\pi} \kcal_{q\bar q g}(\Yqqg)\Bigg\}
\end{equation}
where the LO part is
\begin{equation}
\label{eq:K_qq_LO}
     \kcal_{q \bar q}^\lo(Y_0)  = N_{01}(Y_0)  z_0 (1-z_0)  K_0\left( |\xt_{01}| \overline Q\right) \phi(z_0, \mu_F)
\end{equation}
and the NLO corrections are
\begin{equation}
\label{eq:K_qq_NLO}
\begin{split}
      \kcal_{q \bar q}^\nlo(\Ydip)  =& N_{01}(\Ydip)  z_0 (1-z_0)  K_0\left( |\xt_{01}| \overline Q\right)
        \left\{
        \phi(z_0, \mu_F) \left( \frac{1}{2} \ln^2\left(\frac{z_0}{1-z_0}\right)-\frac{\pi^2}{6}+\frac{5}{2}\right)\right.\\
        &\left.-\ln( \frac{\mu_F^2\xt_{01}^2 e^{2\gamma_E}}{4} )\int_0^1 \dd{z'}  K(z_0,z')\phi(z', \mu_F) 
        \right. \\
        &\left. +\frac{D_s-4}{D-4}
        \int_0^1 \dd{z'} \phi(z', \mu_F)
        \left[-\frac{1}{2} \delta(z_0-z')+\frac{1-z_0}{1-z'} \theta(z_0-z') +\frac{z_0}{z'}\theta(z'-z_0) \right]
    \right\}
\end{split}
\end{equation}
for the dipole part and
\begin{equation}
\label{eq:K_qqg}
\begin{split}
     \kcal_{q \bar q g}(\Yqqg)  =& \frac{2}{\pi z_2} \phi(z_0, \mu_F)\left\{ N_{012}(\Yqqg) K_0(RQ) \frac{1}{1-z_0} \left[
        z_0 \left( (1-z_0-z_2)^2+(1-z_0)^2 \right)\frac{1}{\xt_{21}^2} 
        \right.\right. \\
        &\left. \left.- (1-z_0-z_2) \left( z_0(1-z_0)+ (1-z_0-z_2)(z_0+z_2) \right) \frac{\xt_{20} \vdot \xt_{21}}{\xt_{20}^2 \xt_{21}^2}
    \right]
    \right. \\
    &\left.-N_{01}(\Yqqg) \frac{z_0}{1-z_0} \left( (1-z_0-z_2)^2+(1-z_0)^2 \right) \frac{1}{\xt_{21}^2} e^{-\frac{\xt_{21}^2}{\xt_{01}^2 e^{\gamma_E}}} K_0\left( \xt_{01} \overline Q\right)   \right\}
\end{split}
\end{equation}
for the real emission. The lower limit for the $z_2$ integral is given by Eq.~\eqref{eq:zmin}. This expression is finite and  suitable for numerical evaluation. The rapidity scales at which the different dipole amplitudes are evaluated, $Y_0$, $\Ydip$ and $\Yqqg$, are shown explicitly. In numerical calculations we follow Ref.~\cite{Beuf:2020dxl} and take $Y_0=0$.

The NLO correction to the dipole part has a dependence on the regularization scheme given by a term proportional to
\begin{equation}
    \label{eq:scheme_depedence}
    \frac{D_s-4}{D-4} = 
    \begin{cases}
    1 \textrm{ for CDR} \\
    0 \textrm{ for FDH}.
    \end{cases}
\end{equation}
This regularization scheme dependence is in principle  canceled by the regularization scheme dependence of the distribution amplitude at the given order in $\as$.
The distribution amplitude is a nonperturbative quantity that has to be determined from some process where  the same regularization scheme dependence should also appear. In this paper, we choose to use the CDR regularization scheme when we show numerical results in Sec.~\ref{sec:numerical_results}. However, it will turn out that the regularization scheme dependence is very small even if the same distribution amplitude is used in both schemes, which is a consequence of the fact that for the first term in the Gegenbauer expansion~\eqref{eq:gegenbauer_expansion} of the distribution amplitude this regularization scheme dependent term vanishes.
The regularization scheme dependence of the cross section  will be discussed quantitatively in Appendix~\ref{app:scheme_dependence}.

The dependence on the factorization scale $\mu_F$ is of higher order in $\as$, as can be verified by taking into account the ERBL equation. However, as we are keeping terms only to the order $\as$, the results do have a dependence on the factorization scale. The value of $\mu_F$ can be chosen in different ways. Eq.~\eqref{eq:NLO_qq_fin} suggests the choice $\mu_F^2 = 4e^{-2\gamma_E}/\xt_{01}^2$, as with this choice the logarithm multiplying the ERBL kernel $K(z, z')$ vanishes (we will refer to this term as the ``ERBL term''). Note that the factor $4e^{-2\gamma_E}$ is the same one that appears in the Fourier analysis of the coordinate space running coupling~\cite{Kovchegov:2006vj,Lappi:2012vw}. This choice for the factorization scale will be referred to as the $r$-scheme to emphasize its dependence on the dipole size. In the $q \bar qg$ term we choose to use the smallest dipole size $\min\{|\xt_{01}|,|\xt_{20}|,|\xt_{21}|\}$ for the factorization scale, in accordance with the running of the coupling constant $\as$. 
Another possible choice for the factorization scale is to use $\mu_F = Q$, which is supported by the fact that the relevant length scales for meson production are $Q \sim 1/|\xt_{01}|$, meaning that the logarithm $\ln(Q^2 \xt_{01}^2)$ of the ERBL term in Eq.~\eqref{eq:NLO_qq_fin} should also be small in this scheme. This will be referred to as the $Q$-scheme, and its main advantage is that the hard scale does not depend on the integration variable. In this paper, we will use the $r$-scheme in our calculations as then the ERBL term vanishes completely. In the $Q$-scheme, there can in practice be a large contribution from the ERBL term as the dipole amplitude amplifies the contribution of larger dipoles that can have a numerically significant contribution even at moderately large virtualities~\cite{Mantysaari:2018zdd,Mantysaari:2018nng}. This scheme dependence is studied in more detail in Appendix~\ref{app:scheme_dependence}, where it is shown that the factorization scale dependence at the cross section level is a few percent.

The result \eqref{eq:total_NLO} can be compared to the previously calculated NLO light vector meson production from Ref.~\cite{Boussarie:2016bkq}. In that paper the production amplitude is presented in the momentum space as opposed to the mixed space used in this paper. Comparing these results is non-trivial, as one has to perform complicated Fourier transforms from momentum space to coordinate space to match the results. We have only been able to make comparisons for the dipole part of the amplitude, finding that the result of Ref.~\cite{Boussarie:2016bkq} matches our result in the CDR scheme apart from differences in the UV subtraction procedure. 
The real gluon emission part is much more complicated, and so far we have not been able to make actual comparisons of the results.

\section{Numerical results}
\label{sec:numerical_results}

In this section, we present numerical results for coherent light vector meson electroproduction at next-to-leading order, calculated using Eq.~\eqref{eq:total_NLO}. As our default setup, we use the CDR scheme for regularization and $r$-scheme for the factorization scale $\mu_F$. For the distribution amplitude, we choose to keep only the first two terms in the Gegenbauer expansion~\eqref{eq:gegenbauer_expansion}. The reason for this is that the exact values for the higher-order terms  are not well known, but estimated to be small~\cite{Gao:2014bca}. For the $\rho$ meson, the coefficient of the second term has been extracted in many different ways, with relatively large uncertainties~\cite{Polyakov:2020cnc}. We choose to use the value
$a_2(\mu_F = 1\, \gev) = 0.1$ which is in agreement with most of the values tabulated in Ref.~\cite{Polyakov:2020cnc}. We also choose to use this same value for the $\phi$ meson, as current analyses suggest that they are of the same order of magnitude~\cite{Ball:1998sk,Gao:2014bca}. As we then use the same distribution amplitude for both mesons, the only difference between $\rho$ and $\phi$ production is the decay constant $f_V$ which appears as an overall coefficient in Eq.~\eqref{eq:total_NLO}. These decay constants can be calculated from the experimental values for the leptonic widths~\cite{ParticleDataGroup:2020ssz} using Eq.~\eqref{eq:leptonic_width}. 

The numerical results are calculated using the dipole amplitude fits from Refs.~\cite{Beuf:2020dxl,heikki_mantysaari_2020_4229269} for the different schemes of the BK evolution equation discussed in Sec.~\ref{sec:bkevolution}. We use the fits where the ``Balitsky+smallest dipole'' running coupling scheme is used, and use both fits with initial evolution rapidities $\Ybk, \eta_{0,\mathrm{BK}}=0$ and $\Ybk,\eta_{0,\mathrm{BK}}=4.61$ (in which case the dipole amplitude is frozen in the region $Y_0=0<Y<\Ybk$ or $\eta_0=0<\eta<\eta_{0,\mathrm{BK}}$). In these fits the impact parameter dependence is assumed to factorize and one can replace $\int \dd[2]\bt \to \sigma_0/2$, and the proton transverse area $\sigma_0/2$ is a fit parameter which is also determined in Ref.~\cite{Beuf:2020dxl}.

\begin{figure}
	\centering
    \begin{subfigure}{0.45\textwidth}
        \centering
        \includegraphics[width=\textwidth]{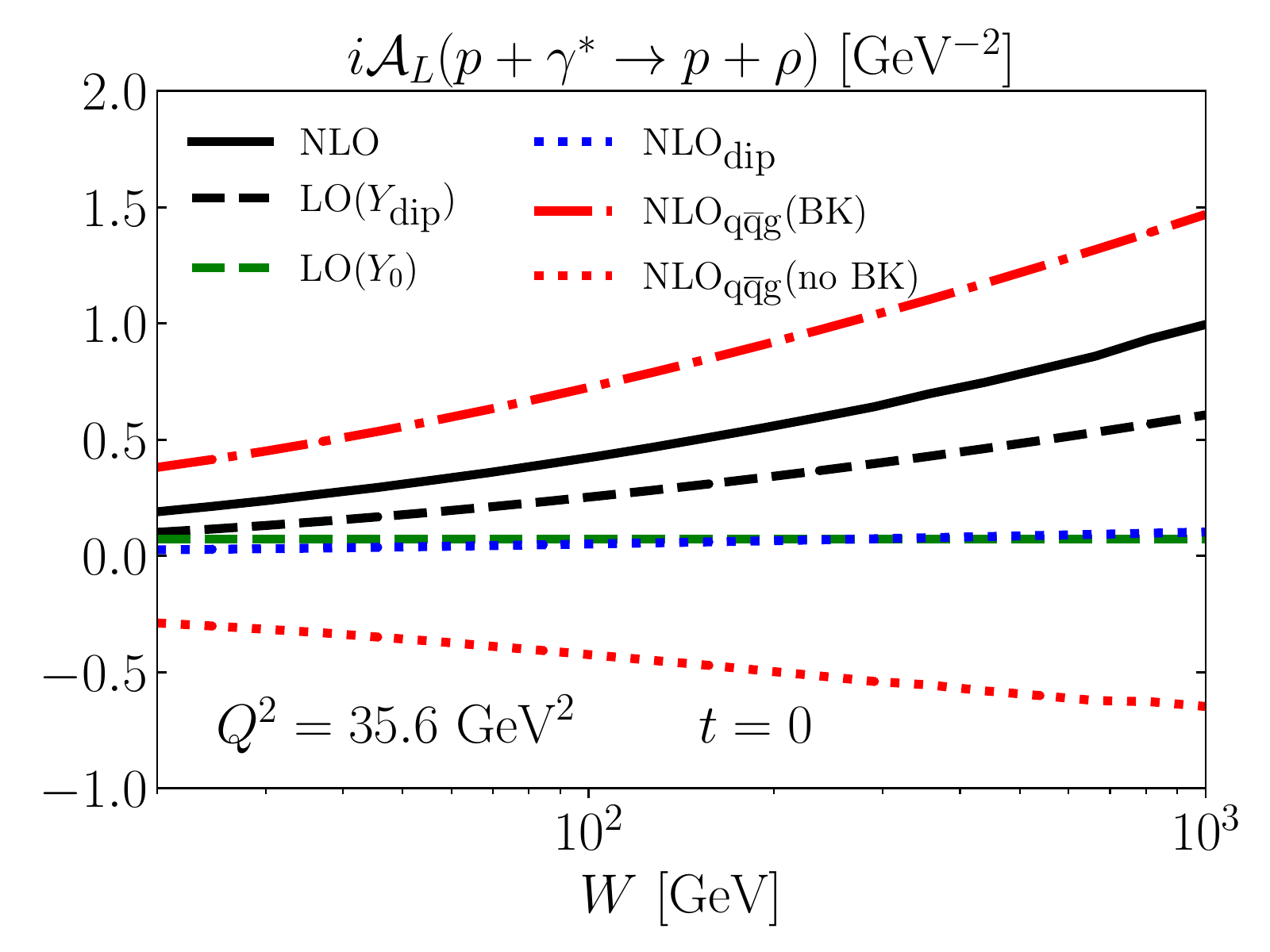}
        \caption{ Dependence on the center-of-mass energy of the $\gamma^*-p$ system. }
        \label{fig:amp_fixed_Q}
    \end{subfigure}
    \begin{subfigure}{0.45\textwidth}
        \centering
        \includegraphics[width=\textwidth]{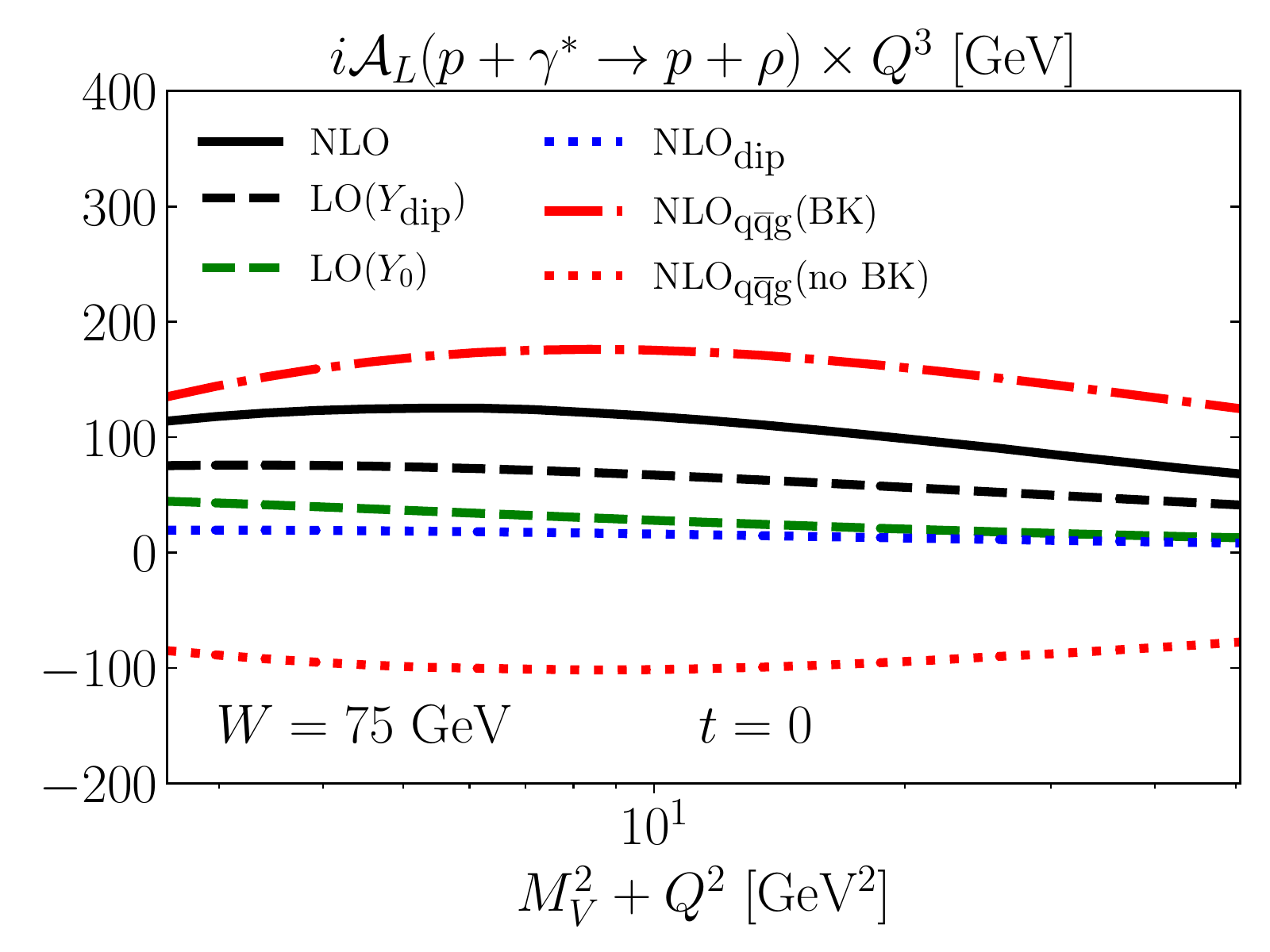}
        \caption{ Dependence on the photon virtuality $Q^2$. The amplitude has been scaled by $Q^3$ for easier readability. }
        \label{fig:amp_fixed_W}
    \end{subfigure}
	  \caption{	 Different parts of the longitudinal NLO amplitude for exclusive $\rho$ production. }
        \label{fig:amp}
\end{figure}

In Fig.~\ref{fig:amp} we show different contributions to the exclusive $\rho$ production amplitude at NLO as a function of the center-of-mass energy $W$ (Fig.~\ref{fig:amp_fixed_Q}) and photon virtuality $Q^2$ (Fig.~\ref{fig:amp_fixed_W}). The same dipole amplitude, corresponding to the KCBK equation with the initial rapidity $\Ybk = 4.61$ for the BK evolution~\cite{Beuf:2020dxl}, is used in these figures.
Here the leading-order result is denoted by $\lo(\Ydip)$ which is calculated from the leading-order part of Eq.~\eqref{eq:total_NLO} with the dipole amplitude evaluated at rapidity $\Ydip$. 
Using the evolved rapidity $\Ydip$ means that the $\lo(\Ydip)$ contains the resummation of large logarithms $\sim \as \ln 1/x$ included in the BK evolution. The result $\lo(Y_0)$ is the leading-order term of Eq.~\eqref{eq:total_NLO} at the initial rapidity $Y_0$, and $\nlo_\text{dip}$ is the NLO correction to the dipole term corresponding to Eq.~\eqref{eq:K_qq_NLO}. The contribution from the $q \bar q g$ term, Eq.~\eqref{eq:K_qqg}, has been divided into two parts: the result $\nlo_{q \bar qg}(\textrm{BK})$ contains only the part corresponding to the BK equation, Eq.~\eqref{eq:qqg_singularity}, and  $\nlo_{q \bar qg}(\textrm{no BK})$ is Eq.~\eqref{eq:K_qqg} from which the BK contribution has been subtracted. The total NLO result is then the sum
\begin{equation}
\label{eq:nloamp_decomposition}
    \nlo = \lo(Y_0) + \nlo_\textrm{dip} + \nlo_{q \bar qg}(\textrm{BK}) +\nlo_{q \bar qg}(\textrm{no BK}). 
\end{equation}

From these plots we see that the contributions from the LO result at the initial rapidity and from the NLO dipole term are small. Both $q \bar q g$ contributions are large, but they mainly cancel each other. These findings are similar to what has been observed in the case of heavy vector meson production~\cite{Mantysaari:2021ryb}. The total NLO correction, the difference between $\nlo$ and $\lo(\Ydip)$, is large and positive. However, we point out that in these plots the same NLO fitted dipole amplitude was used to calculate all of the results. Consequently, these results only quantify the largeness of the NLO correction terms in Eq.~\eqref{eq:total_NLO} and not the actual difference between the NLO and LO results where the corresponding dipole amplitude fits should be used. 

We also note that at fixed coupling the identification $\lo(\Ydip) = \lo(Y_0) +  \nlo_{q \bar qg}(\textrm{BK})$ would be exact if the dipole amplitude satisfied the leading-order fixed coupling BK equation. In that case there would be no ambiguity in defining the leading-order amplitude. In our setup this is not the case, and consequently the leading-order amplitude is not uniquely defined. In this work we choose  it to be $\lo(\Ydip)$ following Ref.~\cite{Mantysaari:2021ryb}, as this is the most natural choice when using a dipole amplitude that satisfies a resummed BK evolution equation. Identifying $\lo(Y_0) +  \nlo_{q \bar qg}(\textrm{BK})$ as a leading order term instead would have maximally a $\sim 20\%$ effect on the calculated cross sections discussed below.

Next we show numerical comparisons to the existing coherent vector meson production data for $\rho$ and $\phi$ mesons at (moderately) large $Q^2$. The H1 data is from Ref.~\cite{H1:2009cml}, and the ZEUS data is from Ref.~\cite{ZEUS:2005bhf} for $\phi$ and Ref.~\cite{ZEUS:1998xpo} for $\rho$. The results are shown with various different dipole amplitude fits that all give a good description of the HERA structure function data. As discussed above, the NLO results use fits from Ref.~\cite{Beuf:2020dxl}. For the leading order, the dipole amplitude used is the ``MV$^e$" fit from Ref.~\cite{Lappi:2013zma}. In the leading-order calculation the evolution rapidity is chosen as $Y=\ln\frac{1}{\xpom}=\ln\frac{W^2+Q^2}{Q^2+M_V^2}$, consistently with the fit. 
The differential cross section is proportional to the square of the production amplitude as given by Eq.~\eqref{eq:dsigma_dt}. When calculating the cross section at NLO, we drop the higher-order terms  proportional to $\as^2$ so that we only keep the genuine NLO correction at the cross section level. 

\begin{figure}
	\centering
    \begin{subfigure}{0.45\textwidth}
        \centering
        \includegraphics[width=\textwidth]{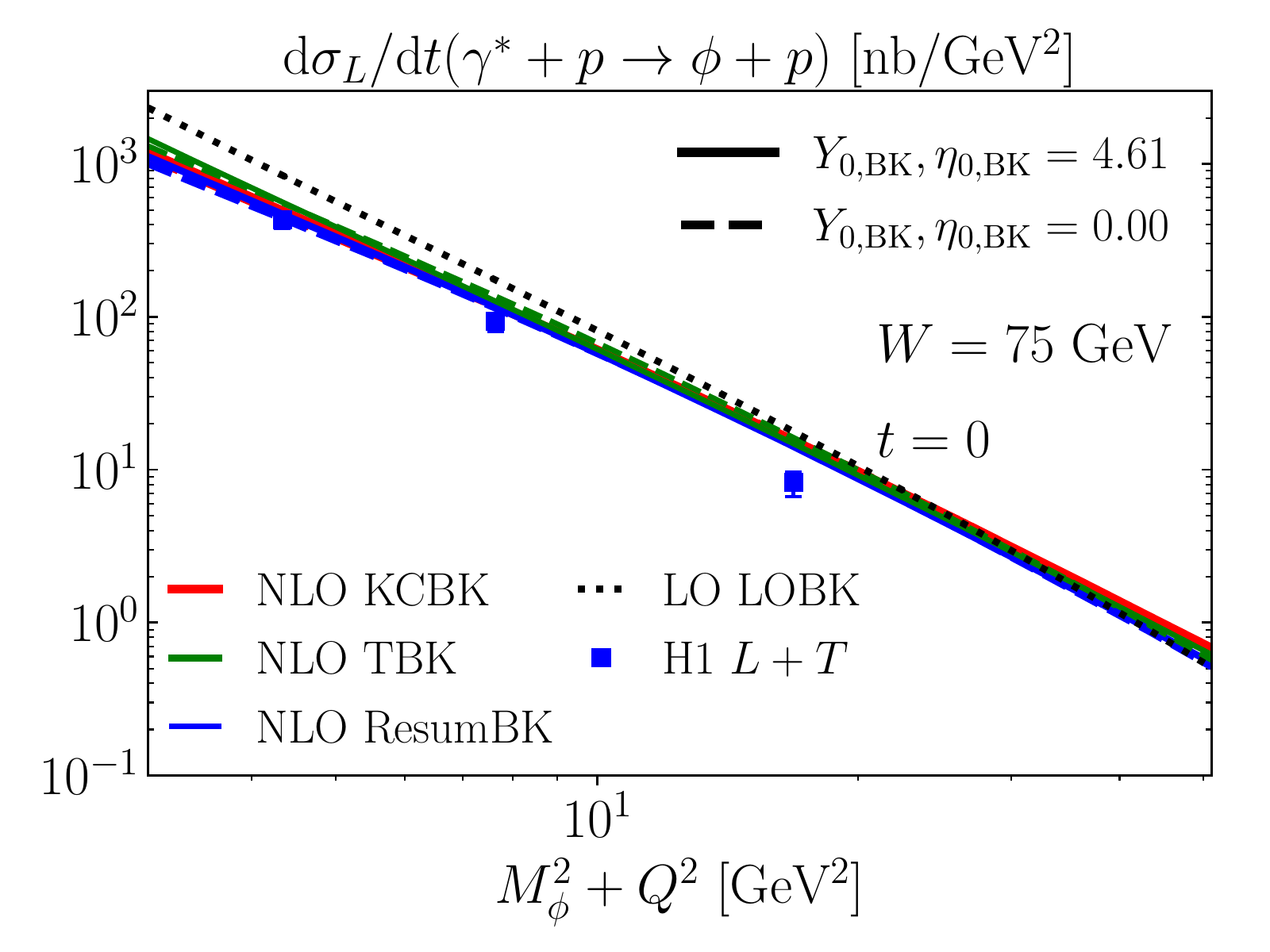}
         \caption{Cross section for $\phi$ production.
         }
        \label{fig:diff_cs_phi_fixed_W}
    \end{subfigure}
    \begin{subfigure}{0.45\textwidth}
        \centering
        \includegraphics[width=\textwidth]{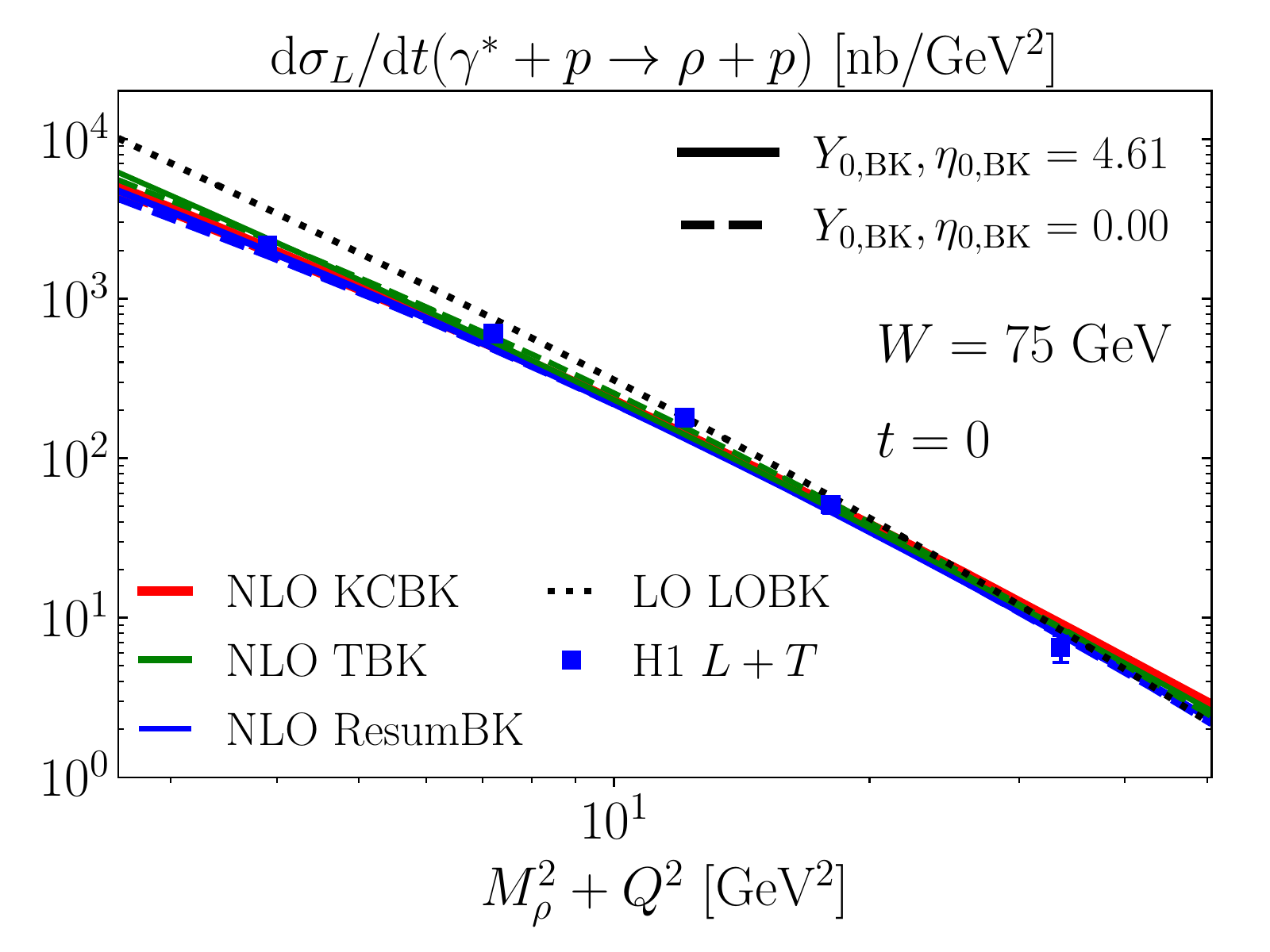}
        \caption{ Cross section for $\rho$ production.}
        \label{fig:diff_cs_rho_fixed_W}
    \end{subfigure}
	  \caption{	Photon virtuality dependence of the longitudinal  cross section at $t=0$ for various different dipole amplitude fits, compared to the H1 data for the sum of longitudinal and transverse productions~\cite{H1:2009cml}. }
        \label{fig:diff_cs}
\end{figure}

In Fig.~\ref{fig:diff_cs} we show the differential cross section for the longitudinal $\phi$ and $\rho$ production at $t=0$. Here the experimental data is for the total production which is the sum of the longitudinal and transverse channels. However, the longitudinal production dominates at $Q^2 \gg M_V^2$ and therefore it is expected that for high virtualities these data points accurately correspond to the longitudinal case. 

In general, we see that both the LO and NLO results describe the H1 data well. 
The difference between the LO and NLO results is smaller than one would expect based on Fig.~\ref{fig:amp}, as in the leading-order fit the nonperturbative parameters describing the initial condition of the dipole amplitude effectively capture part of the higher-order effects. 
This difference becomes  small 
at high virtualities, where our approach is expected to be most reliable. The NLO results also give a surprisingly accurate description of the data for smaller values of the photon virtuality where the framework cannot be trusted (we have assumed that $Q^2\gg M_V^2$). 
We consider this agreement to be accidental for two reasons. First, we have only calculated the longitudinal cross section, leaving out the transverse contribution which is significant at small $Q^2$. Secondly,  we neglect any dependence on the dipole size in the meson wave function, keeping only the $\rt=0$ case which corresponds to the distribution amplitude. In general, the wave function is expected to be a decreasing function of $|\rt|$, meaning that this approximation overestimates the results. These two corrections, which may have significant numerical contributions at small virtualities, affect the result in opposite ways and therefore their total contribution at least partially cancels. 

\begin{figure}
    \centering
    \includegraphics[width=0.45\textwidth]{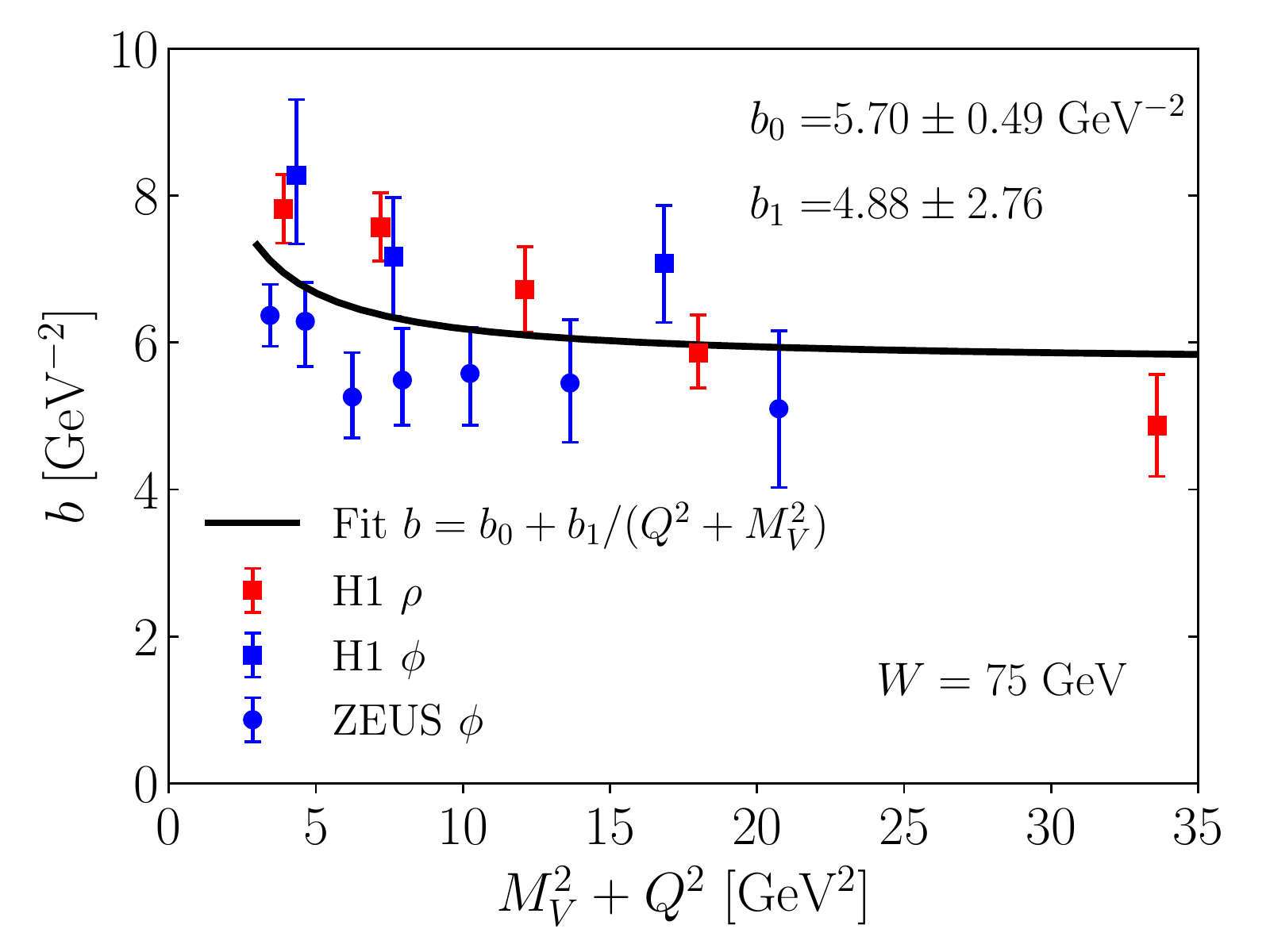}
	\caption{The measured slope parameter $b$ for $\rho$ and $\phi$ production as a function of photon virtuality~\cite{H1:2009cml, ZEUS:2005bhf, ZEUS:1998xpo} and a fit to this data. }
    \label{fig:b_slope_fit}
\end{figure}

Next we will consider the $t$-integrated cross sections for which more data exists. To avoid additional modeling for the impact parameter dependence of the dipole amplitude, we evaluate the $t$-integral by using the following experimental parametrization for the $t$-dependence of the cross section:
\begin{equation}
    \label{eq:xs_t-dependence}
    \frac{\dd{\sigma}}{\dd{t}} = e^{-b|t|} \times \frac{\dd{\sigma}}{\dd{t}}(t=0).
\end{equation}
Here $b$ is the slope parameter that in general depends on $Q^2$ and $W$. It has been measured for both $\phi$ and $\rho$~\cite{H1:2009cml, ZEUS:2005bhf, ZEUS:1998xpo} at different values of the virtuality at $W=75\,\gev$. The slope parameter can be thought of as the effective transverse area of the meson-target system, and we model its dependence on virtuality and center-of-mass energy by assuming the parametrization
\begin{equation} 
b=b_0 + \frac{b_1}{Q^2+M_V^2} +4 \alpha' \ln \frac{W}{W_0}.
\end{equation} 
The $W$ dependence determined from HERA data~\cite{H1:2009cml} gives $\alpha'=0.12 \pm 0.04$. The model for the virtuality dependence is chosen for its simplicity and that it approaches a constant value at high $Q^2$. Also, the dependence on the virtuality and the center-of-mass energy does not seem to be correlated~\cite{H1:2009cml}. We fit the parameters $b_0$ and $b_1$ to H1 and ZEUS data at $W_0=75 \,\gev$, with the fit shown in Fig.~\ref{fig:b_slope_fit}, and
note that the errors on these fitted parameters are significant,
which results in $\sim 10\%$ uncertainty in the calculated total cross sections. 

\begin{figure}
	\centering
    \begin{subfigure}{0.45\textwidth}
        \centering
        \includegraphics[width=\textwidth]{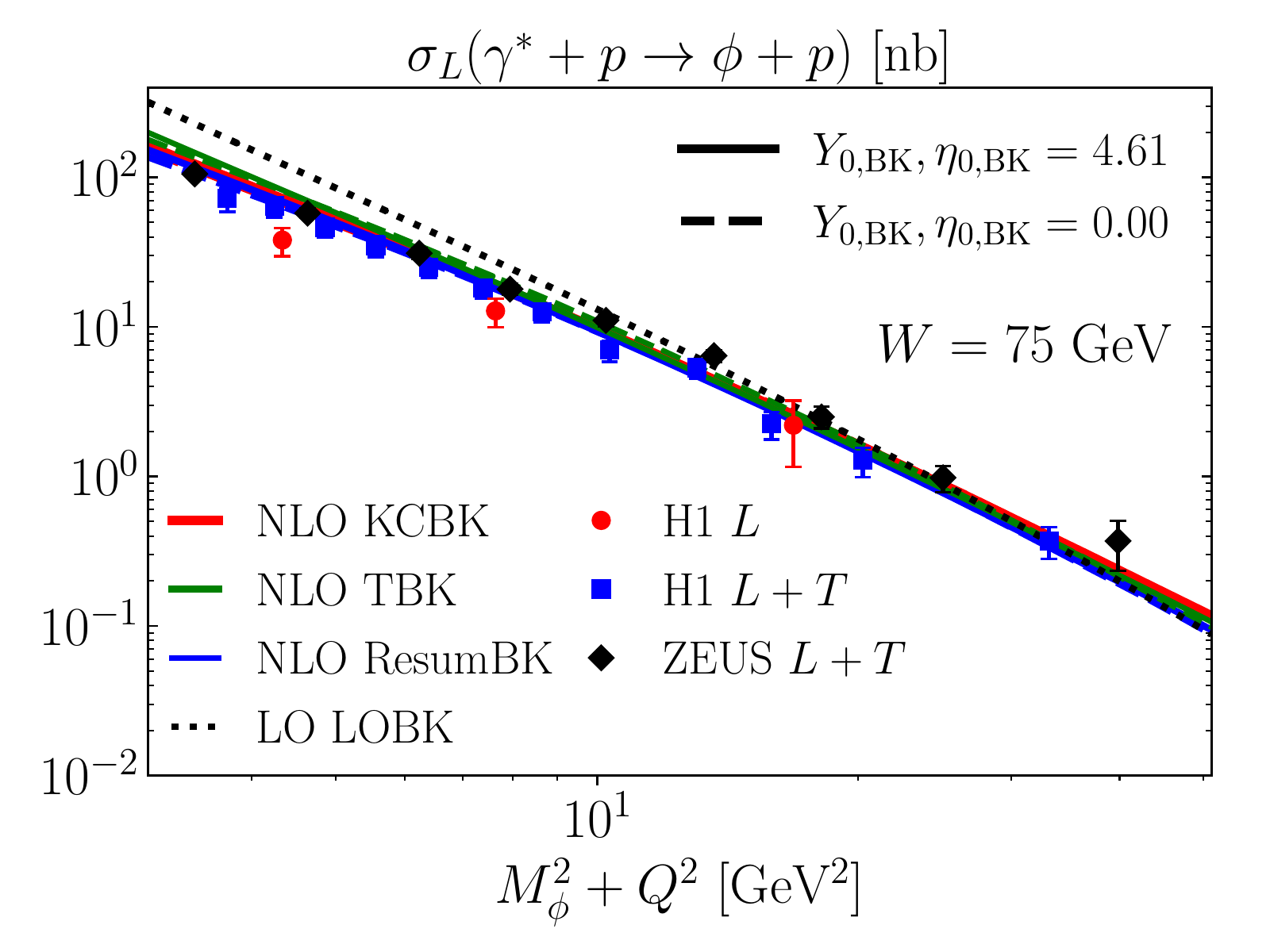}
        \caption{ Cross section for $\phi$ production. }
        \label{fig:cs_phi_fixed_W}
    \end{subfigure}
    \begin{subfigure}{0.45\textwidth}
        \centering
        \includegraphics[width=\textwidth]{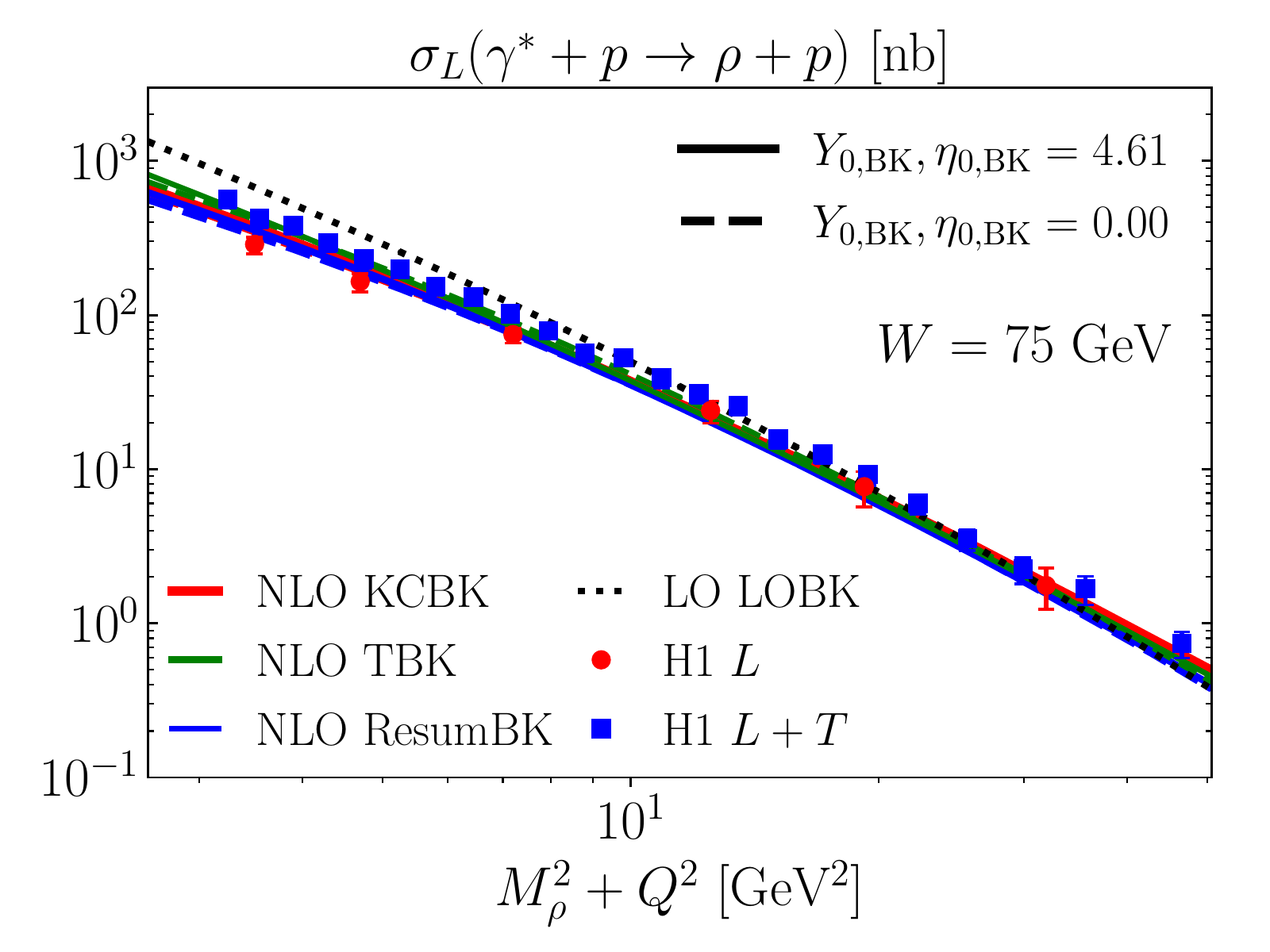}
        \caption{ Cross section for $\rho$ production. }
        \label{fig:cs_rho_fixed_W}
    \end{subfigure}
	  \caption{	Photon virtuality dependence of the integrated longitudinal cross section for various different dipole amplitude fits, compared to the HERA data~\cite{H1:2009cml, ZEUS:2005bhf, ZEUS:1998xpo}.   }
        \label{fig:integrated_cs_W}
\end{figure}

\begin{figure}
	\centering
    \begin{subfigure}{0.45\textwidth}
        \centering
        \includegraphics[width=\textwidth]{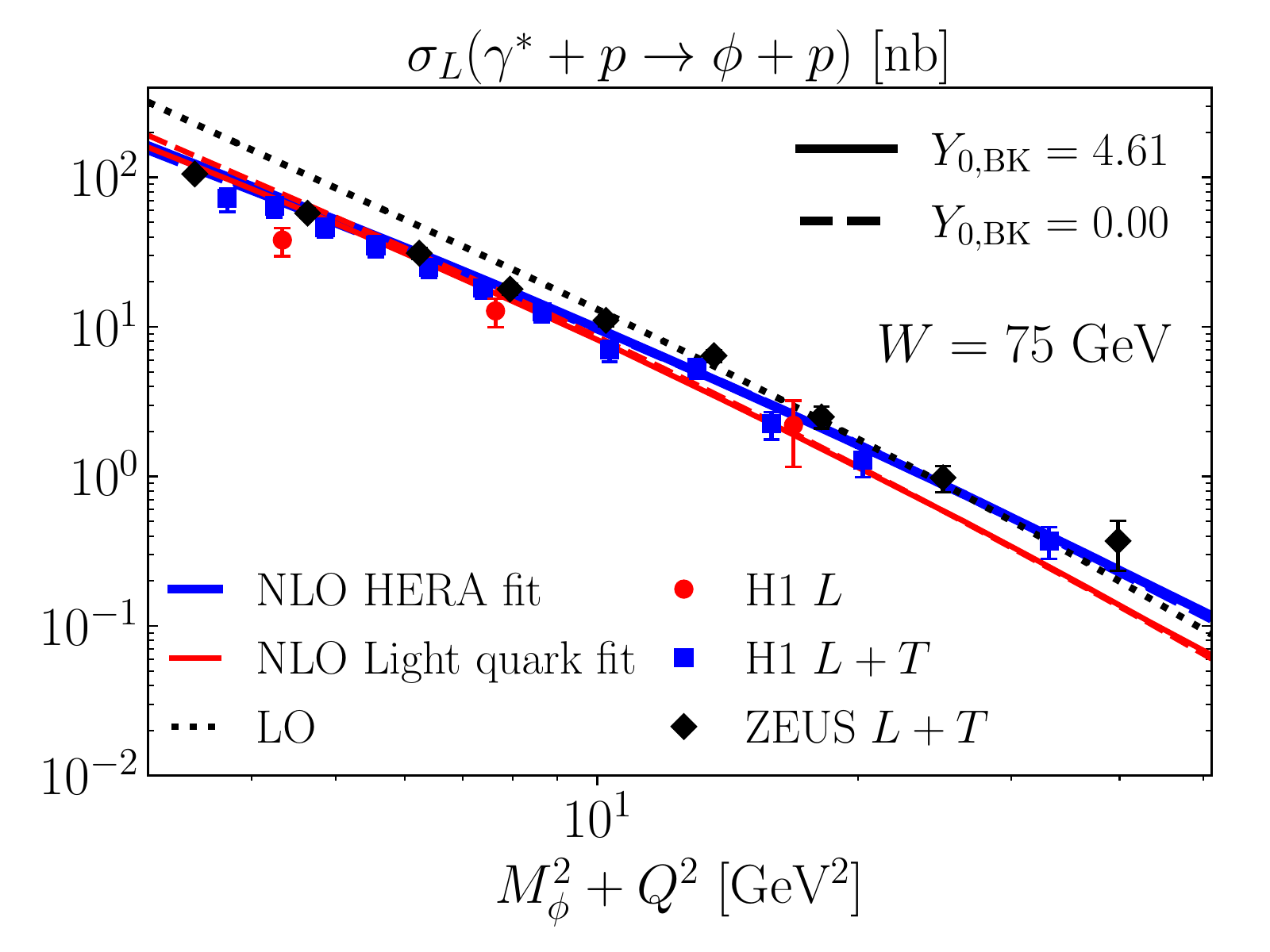}
        \caption{ Cross section for $\phi$ production. }
        \label{fig:phi_HERA_vs_lightq_fixed_W}
    \end{subfigure}
    \begin{subfigure}{0.45\textwidth}
        \centering
        \includegraphics[width=\textwidth]{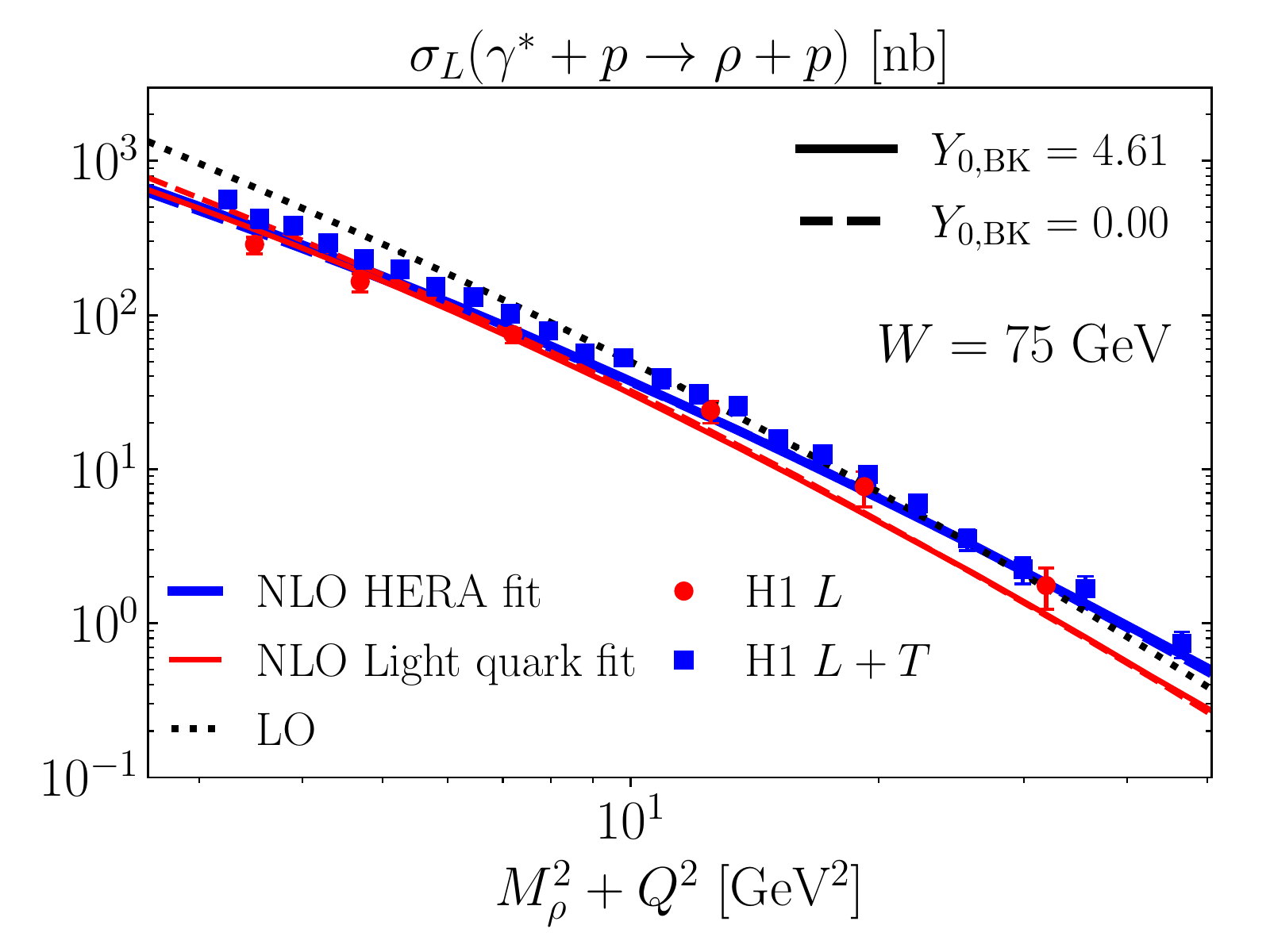}
        \caption{ Cross section for $\rho$ production. }
        \label{fig:rho_HERA_vs_lightq_fixed_W}
    \end{subfigure}
	  \caption{	Photon virtuality dependence of the integrated cross section for longitudinal production, with dipole amplitudes fitted to HERA structure function data and pseudodata consisting of only light quark production.  }
        \label{fig:HERA_vs_lightq_W}
\end{figure}

The virtuality dependence of the coherent $\phi$ and $\rho$ production cross sections is shown in Figs.~\ref{fig:integrated_cs_W} and \ref{fig:HERA_vs_lightq_W}. In Fig.~\ref{fig:integrated_cs_W}, the results are calculated using different dipole amplitudes fitted to the HERA structure function data in Ref.~\cite{Beuf:2020dxl}, using fits with both choices for the initial evolution rapidities $\Ybk$ ($\eta_{0,\mathrm{BK}}$ in the case of TBK evolution). The H1 collaboration has measured, in addition to the total production cross section, the longitudinally polarized $\rho$ production, which exactly corresponds to the presented theory calculations. In general we find an excellent agreement with the H1 and ZEUS data~\cite{H1:2009cml, ZEUS:2005bhf, ZEUS:1998xpo}, except that the $\phi$ production cross section is overestimated at low virtualities where our approximations are not justified.

In Fig.~\ref{fig:HERA_vs_lightq_W}, we  show results obtained using the dipole amplitudes fitted to the structure function pseudodata generated in Ref.~\cite{Beuf:2020dxl} such that it includes only the approximative light quark contribution. For comparison, the results calculated with the dipole amplitudes fitted to the full HERA structure function data are also shown.
In the fit process of Ref.~\cite{Beuf:2020dxl} only the light quark contribution is calculated, and as such the fit to light quark pseudodata is in principle better motivated than the fit to the full HERA structure function data. On the other hand, the light-quark-only data contains a larger nonperturbative contribution, and the determined parametrizations describing the initial condition of the dipole amplitude are not physically as well motivated. Here we use these light quark fits with the KCBK evolution equation, but different schemes for the BK evolution result in very similar cross sections at all $Q^2$.

 We see that the results calculated with dipole amplitudes fitted to the light quark pseudodata also show a relatively good agreement with the virtuality dependence of the H1 and ZEUS light meson production data. However, the cross sections at large virtualities are somewhat underestimated. 
This difference in $Q^2$ dependence between the two fit setups is expected, as the light-quark-only pseudodata is close to the full structure function data at low $Q^2$ where similar results for other observables are also expected. On the other hand, at high $Q^2$ the charm contribution on structure functions is significant, and consequently the light quark fit should result in smaller cross sections in this kinematical region, which is exactly what we observe in Fig.~\ref{fig:HERA_vs_lightq_W}.

\begin{figure}
	\centering
    \begin{subfigure}{0.45\textwidth}
        \centering
        \includegraphics[width=\textwidth]{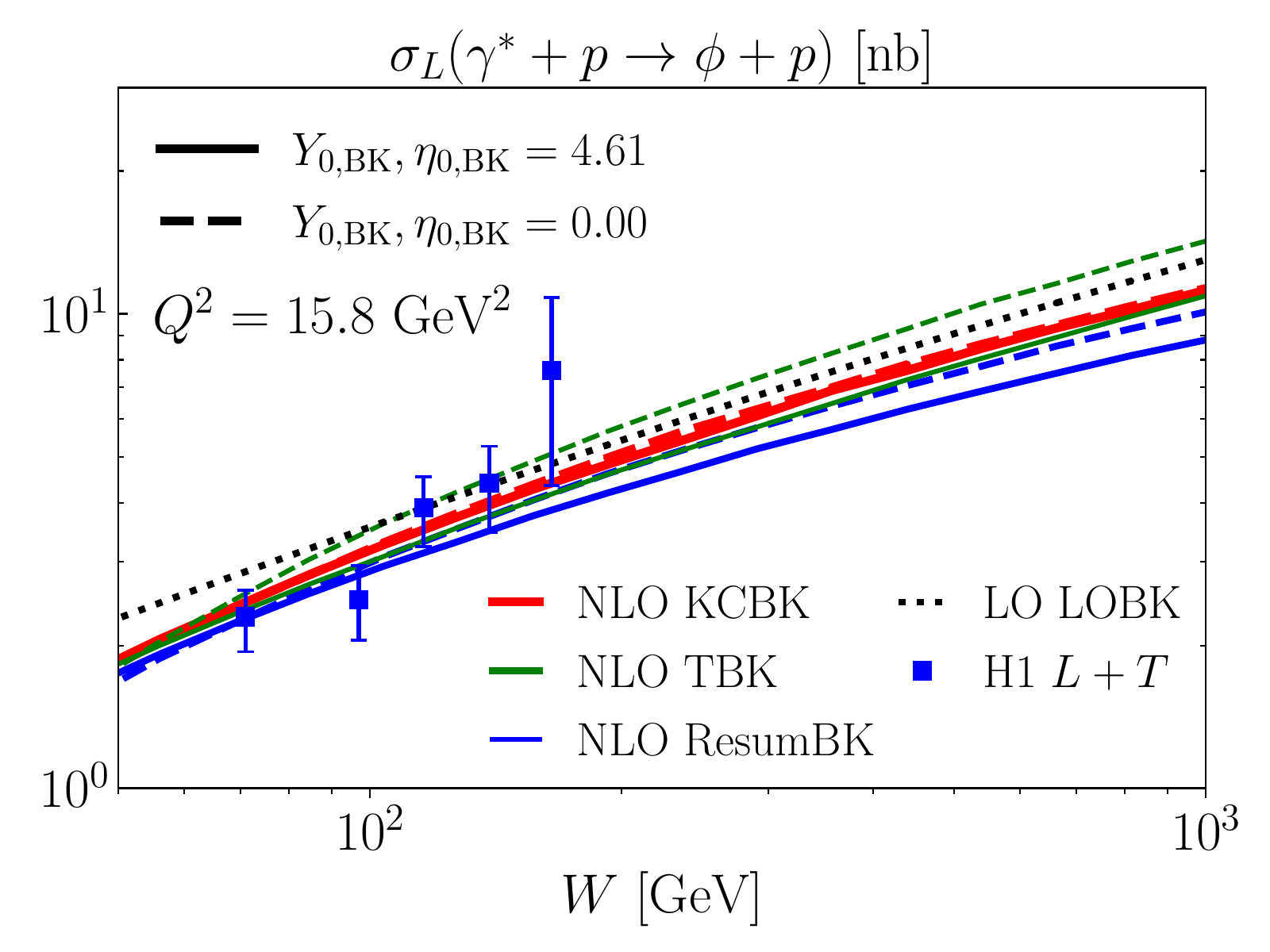}
        \caption{ Cross section for $\phi$ production. }
        \label{fig:cs_phi_fixed_Q}
    \end{subfigure}
    \begin{subfigure}{0.45\textwidth}
        \centering
        \includegraphics[width=\textwidth]{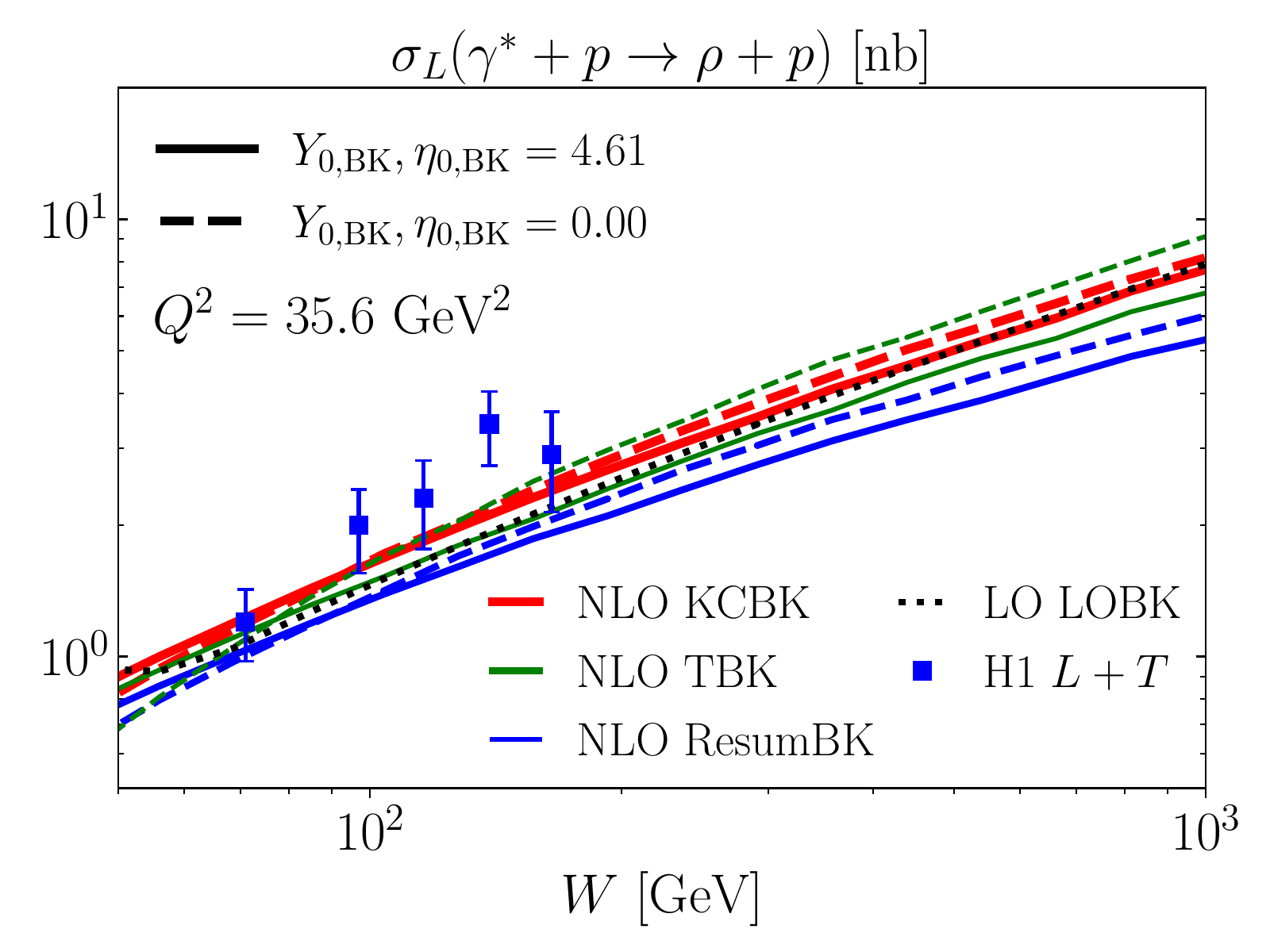}
        \caption{ Cross section for $\rho$ production. }
        \label{fig:cs_rho_fixed_Q}
    \end{subfigure}
	  \caption{	Center-of-mass dependence of the integrated longitudinal cross section compared to the H1 data~\cite{H1:2009cml}.  }
        \label{fig:integrated_cs_Q}
\end{figure}

\begin{figure}
	\centering
    \begin{subfigure}{0.45\textwidth}
        \centering
        \includegraphics[width=\textwidth]{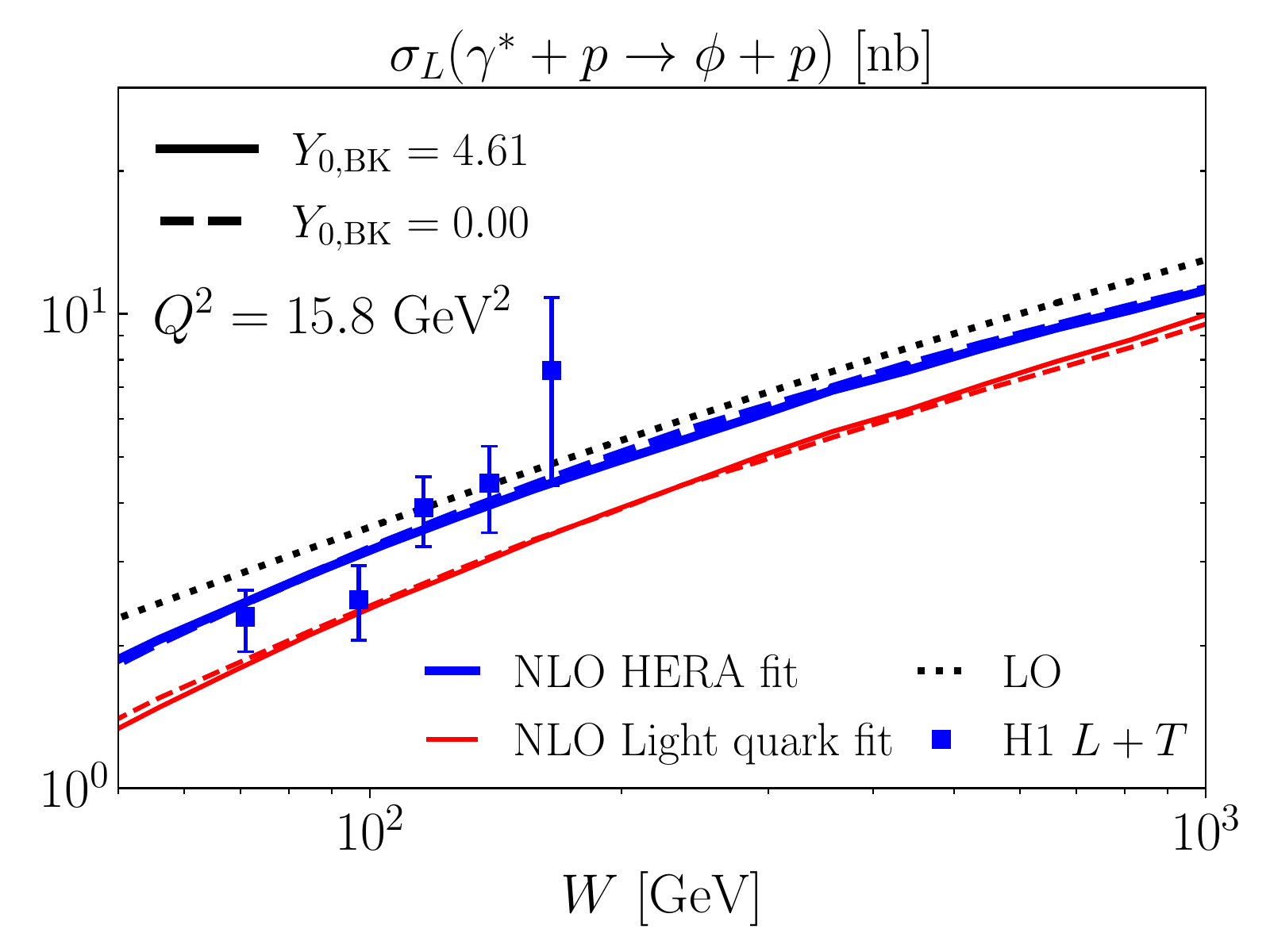}
        \caption{ Cross section for $\phi$ production.}
        \label{fig:phi_HERA_vs_lightq_fixed_Q}
    \end{subfigure}
    \begin{subfigure}{0.45\textwidth}
        \centering
        \includegraphics[width=\textwidth]{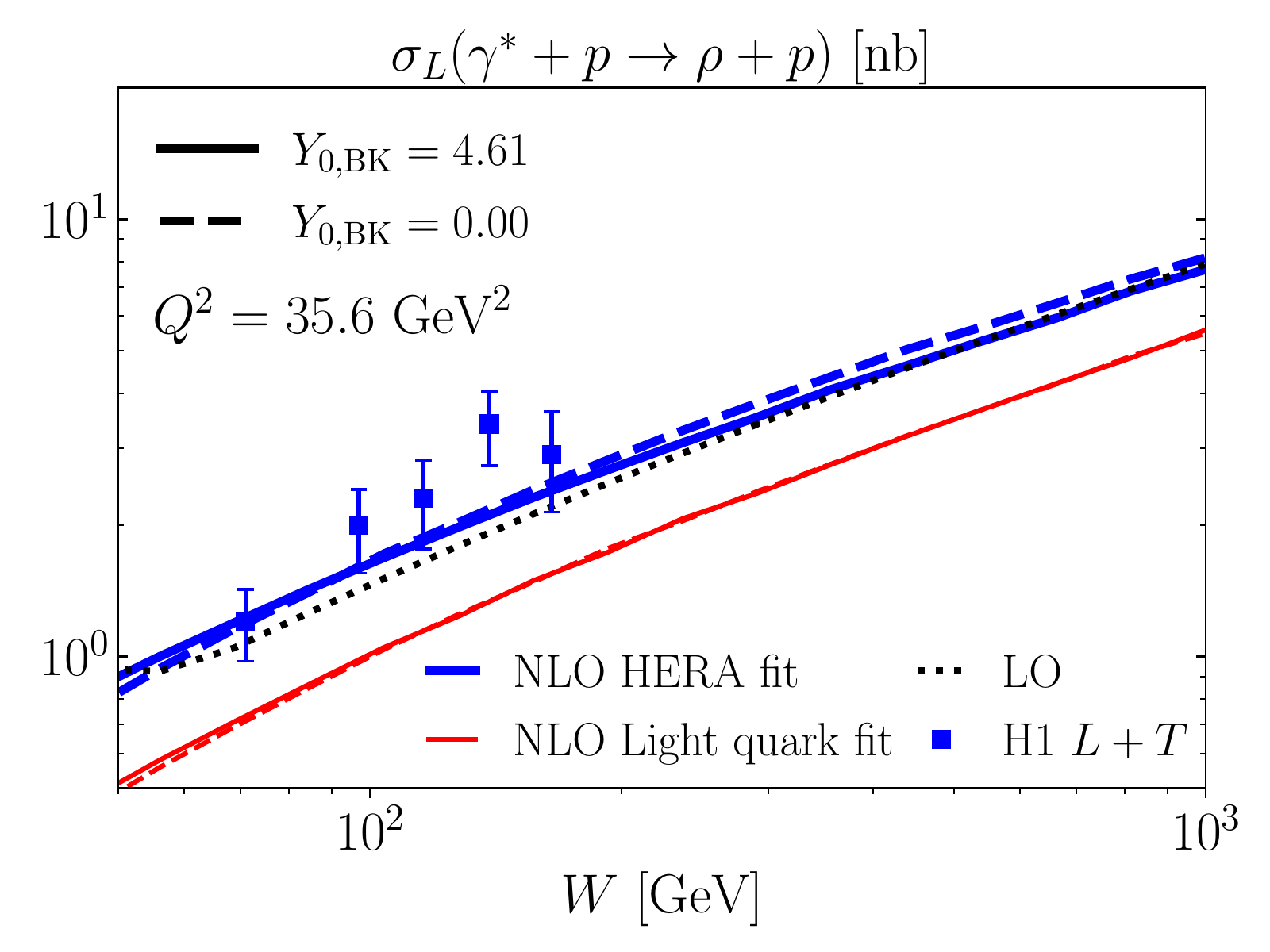}
        \caption{ Cross section for $\rho$ production. }
        \label{fig:rho_HERA_vs_lightq_fixed_Q}
    \end{subfigure}
	  \caption{	Center-of-mass dependence of the integrated longitudinal cross section, with dipole amplitudes fitted to HERA structure function data and pseudodata consisting of only light quark production.}
        \label{fig:HERA_vs_lightq_Q}
\end{figure}

In Figs.~\ref{fig:integrated_cs_Q} and \ref{fig:HERA_vs_lightq_Q}, we show the dependence of the integrated cross section on the photon-proton center-of-mass energy $W$. Again, Fig.~\ref{fig:integrated_cs_Q} shows  results obtained with the dipole amplitudes fitted to the HERA data, and Fig.~\ref{fig:HERA_vs_lightq_Q}  shows results calculated with dipole amplitudes fitted to the light quark pseudodata for comparison. The center-of-mass energy dependence of the results agrees with the data, although the results with the light quark fit seem to underestimate the data by a constant factor as already seen in Fig.~\ref{fig:HERA_vs_lightq_W}. The differences in the results with different schemes for the BK evolution start growing at larger $W$, as at high $W$ one starts to be sensitive to the region not constrained by the structure function data. 
This suggests that light meson production data can provide additional constraints when the nonperturbative initial condition for the BK evolution is extracted from experimental data.
The dependence on the center-of-mass energy is similar for the HERA and light quark fitted dipole amplitudes. 

\begin{figure}
	\centering
    \begin{subfigure}{0.45\textwidth}
        \centering
        \includegraphics[width=\textwidth]{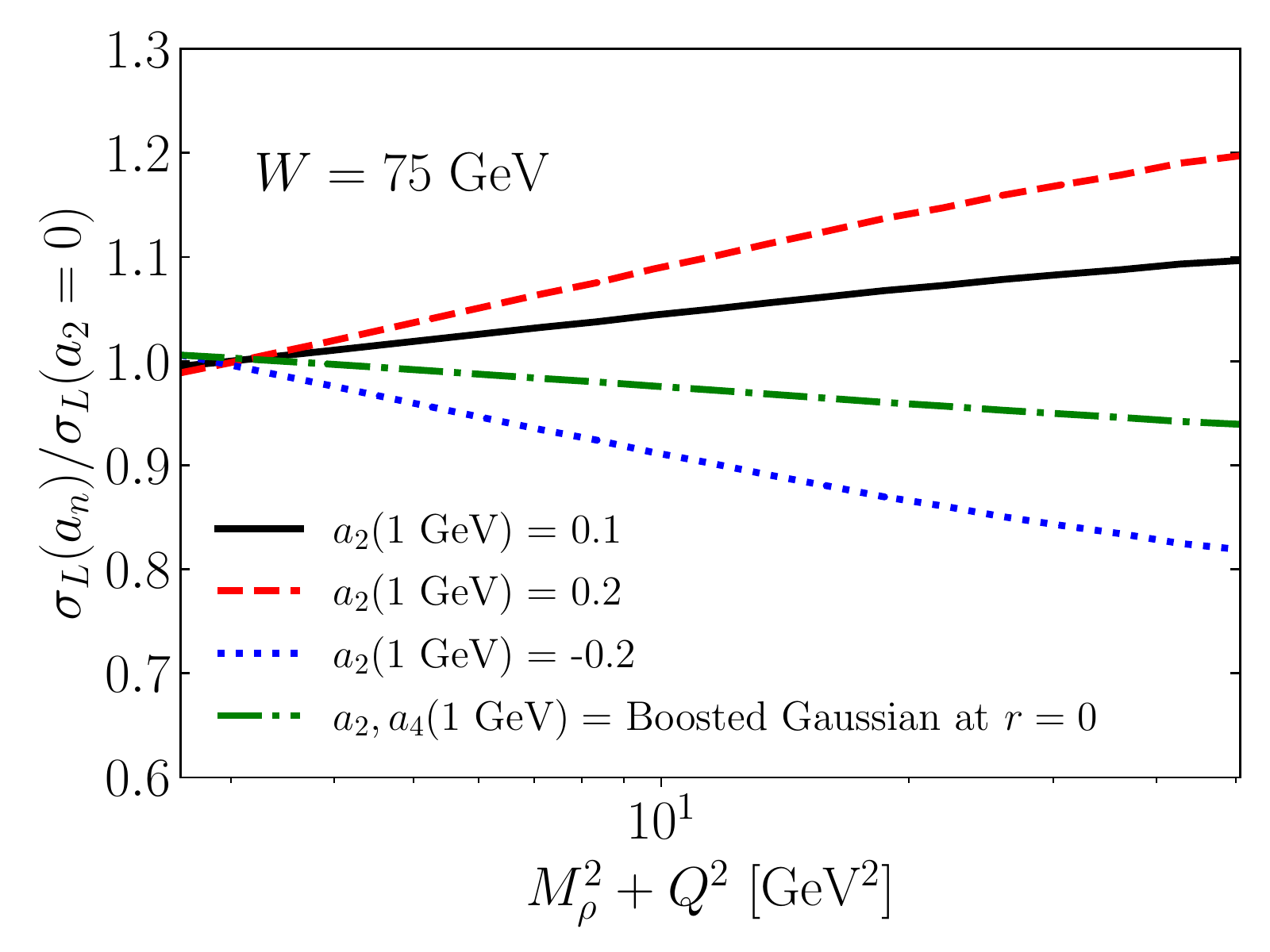}
        \caption{ Ratio of the cross section with varied coefficients in the Gegenbauer expansion to the default setup. 
        }
        \label{fig:gegenbauer_contribution}
    \end{subfigure}
    \begin{subfigure}{0.45\textwidth}
        \centering
        \includegraphics[width=\textwidth]{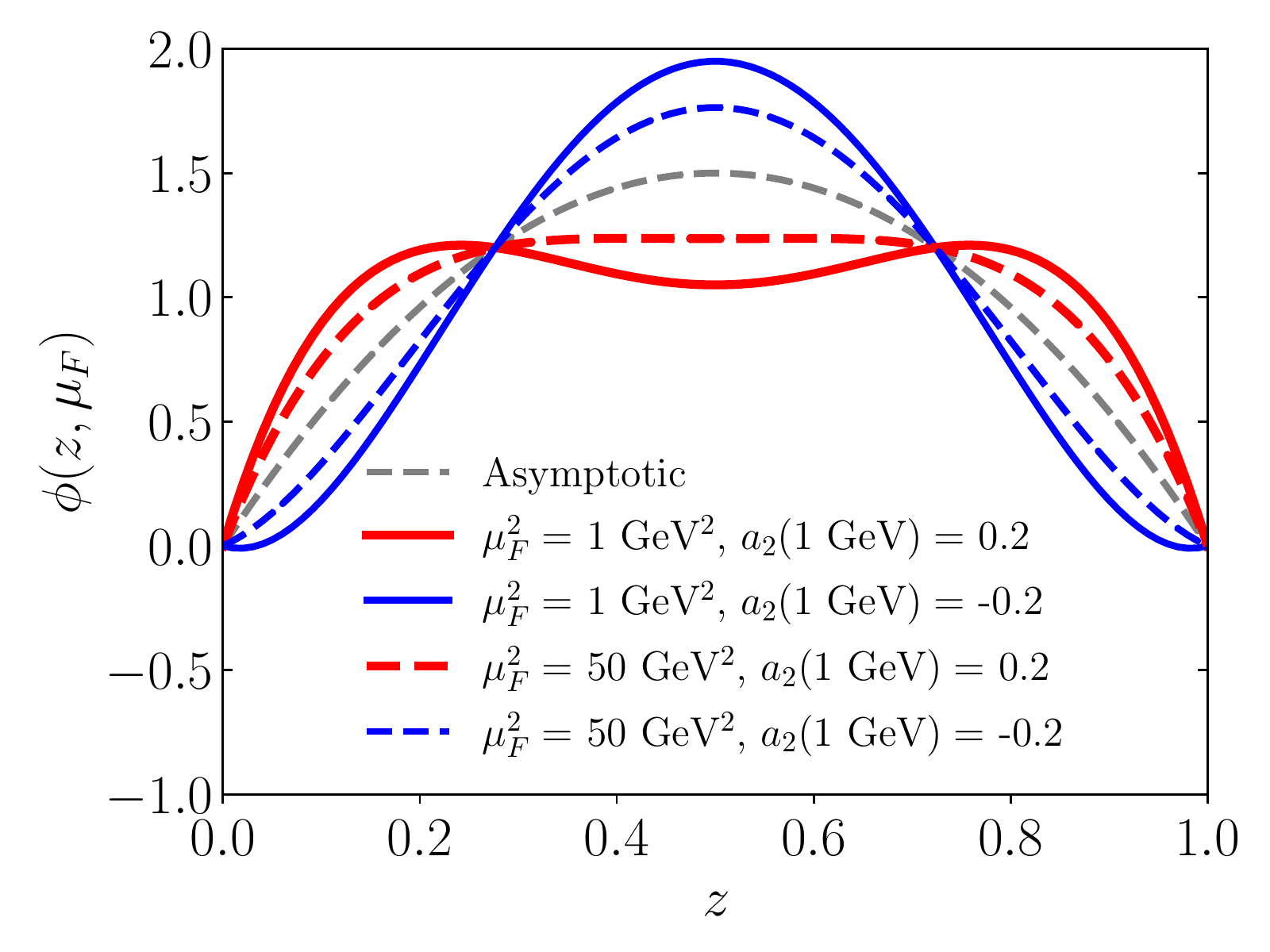}
        \caption{ Distribution amplitude with different coefficients $a_2$ and the effect of the ERBL evolution. The asymptotic form with $a_{n>0}=0$ is also shown. %
        }
        \label{fig:distribution_amplitudes}
    \end{subfigure}
	  \caption{Dependence of the cross section on the distribution amplitude. }
        \label{fig:DA_dependence}
\end{figure}

In Fig.~\ref{fig:gegenbauer_contribution} we show the dependence of the cross section on the distribution amplitude. We have normalized the results by the cross section calculated using the asymptotic form for the distribution amplitude which corresponds to the case where the higher-order terms in the Gegenbauer expansion \eqref{eq:gegenbauer_expansion} vanish, i.e. $a_n=0$ for $n>0$. The case $a_2(1\, \gev) = 0.1$ corresponds to our default setup, and the cases $a_2(1 \, \gev) = \pm 0.2$ are estimates for the upper and lower bounds for the coefficient, chosen based on Ref.~\cite{Polyakov:2020cnc}. The final setup shown has the coefficients $a_2(1 \, \gev)=-0.054$ and $a_4(1 \, \gev)=-0.022$ chosen such that the distribution amplitude matches the Boosted Gaussian wave function parametrization for the $\rho$ meson from Ref.~\cite{Kowalski:2006hc} at $\rt = 0$ where the higher Gegenbauer terms are neglected. The reason for this choice is that the distribution amplitude should roughly correspond to the wave function at $\rt =0$, and the Boosted Gaussian is a phenomenological wave function that describes well vector meson production at leading order~\cite{Kowalski:2006hc}. 
We see that the dependence on the distribution amplitude is moderate, and maximally $\sim 30\%$ in the considered kinematical domain.

The  different  distribution amplitudes are illustrated in 
Fig.~\ref{fig:distribution_amplitudes}, both at the initial scale $\mu_F^2=1\,\gev^2$ and after the ERBL evolution up to $\mu_F^2=50\,\gev^2$ using the extreme values for $a_2$. While the form of the distribution amplitude depends considerably on the value of $a_2$, the effect on the cross section in Fig.~\ref{fig:gegenbauer_contribution} is small.
In Fig.~\ref{fig:distribution_amplitudes} we also see that as the factorization scale $\mu_F$ increases the distribution amplitude approaches the asymptotic form, but there is still a significant deviation from the asymptotic shape at $\mu_F^2=50\,\gev^2$.

\section{Conclusions}
\label{sec:conclusions}
In this paper, we have presented a next-to-leading order calculation of exclusive light vector meson production in the dipole picture using the light-cone perturbation theory.
The main result of this work is the scattering amplitude for longitudinally polarized light vector meson production at high $Q^2$ given in Eq.~\eqref{eq:total_NLO}.
This amplitude is finite and directly suitable for numerical evaluations. In particular the $\frac{1}{D-4}$ divergences have been canceled between the real and virtual diagrams after the  ERBL evolution of the renormalized distribution amplitude is also taken into account. 
The apparent soft gluon divergence is shown to be factorizable into the small-$x$ Balitsky-Kovchegov evolution of the dipole amplitude.
While a similar calculation has already been done in the momentum space~\cite{Boussarie:2016bkq} (using also a different scheme to subtract the rapidity divergence), the results presented in this paper are in the mixed transverse coordinate-longitudinal momentum fraction space where the numerical calculations using existing results for the dipole-target scattering amplitude are straightforward. Comparing the results in the different  spaces is cumbersome due to the requirement of calculating complicated Fourier transforms, and thus an explicit comparison of the two results has been made only partially.

We have also calculated numerically exclusive light meson production at NLO and compared the results  to the existing HERA data for $\rho$ and $\phi$ mesons. The NLO corrections are numerically important, but their effect can be partially captured when the initial condition for the small-$x$ evolution of the dipole amplitude is fitted to the structure function data. Consequently, the differences between LO and NLO results are moderate in the high virtuality region $Q^2  \gg M_V^2$ where our framework is valid. The different 
schemes used to capture higher-order effects in the small-$x$ evolution 
result in similar cross sections for vector meson production. Some deviations can be seen in the center-of-mass energy $W$ dependence, which means that the exclusive vector meson production data can further constrain the nonperturbative initial condition for the small-$x$ evolution. Both the $Q^2$ and $W$ dependencies of the production cross section are in excellent agreement with the HERA data. 
If a dipole amplitude with an initial condition fitted to the structure function pseudodata that only includes a light quark contribution is used, the experimental cross sections are underestimated at high $Q^2$. 
We also note that there is some overall normalization uncertainty due to e.g. modeling the $t$ dependence of the vector meson production cross section.
We additionally left out the commonly used phenomenological corrections (see e.g.~\cite{Kowalski:2006hc}) whose role should be further clarified.

Our result for the analytic expression of the production amplitude is presented in two different schemes for regularization in the transverse plane: the CDR and FDH schemes. The regularization scheme dependence is shown to be very small. The results also depend on the choice for the factorization scale $\mu_F$, for which we present two different choices, taking this scale to be either a function of the dipole size or of the photon virtuality. The dependence on the factorization scale is also relatively small. Both of these scheme dependencies have numerically small effects because the distribution amplitudes for the $\rho$ and $\phi$ mesons are close to the asymptotic form, and the dependence on regularization scheme and factorization scale vanishes in the $Q^2 \rightarrow \infty$ limit. The dependence on the exact form of the distribution amplitude, on the other hand, is somewhat larger with effects of up to $\sim 30\%$ in the HERA kinematics at the cross section level for realistic values of the higher-order terms in the Gegenbauer expansion.

The results in this paper are calculated at zero momentum transfer $t=0$. Calculating the $t$-dependence of exclusive vector meson production is also interesting as it allows access to the spatial distribution of the target color field including its event-by-event fluctuations~\cite{Mantysaari:2020axf,Mantysaari:2016jaz,Mantysaari:2016ykx}. This requires additional nonperturbative modeling of the dipole amplitude which we wanted to avoid in this paper. We also note that the dipole amplitudes used in numerical calculations in this paper were fitted to HERA data using only massless quarks, while there is a significant contribution from the massive $c$ quark to the structure functions in HERA kinematics. As the NLO photon wave functions with massive quarks are becoming available~\cite{Beuf:2021qqa, Beuf:2021srj}, it will be possible to make a new NLO fit for the dipole amplitude to the HERA data including heavy quark contributions. This is needed for accurate phenomenological comparisons with the HERA data. The results of this work can then be used for predicting exclusive light vector meson production in the future EIC which will also produce data for DIS off heavy nuclei, allowing for precision studies of saturation phenomena.

\section*{Acknowledgements}

We thank R. Boussarie, T. Lappi and R. Paatelainen for useful discussions.
This work was supported by the Academy of Finland, the Centre of Excellence in Quark Matter, and projects 338263, 346567	and 321840, by the Finnish Cultural Foundation (J.P), and under the European Union’s Horizon 2020 research and innovation programme by the European Research Council (ERC, grant agreement No. ERC-2018-ADG-835105 YoctoLHC) and by the STRONG-2020 project (grant agreement No 824093).  The content of this article does not reflect the official opinion of the European Union and responsibility for the information and views expressed therein lies entirely with the authors.

\bibliographystyle{JHEP-2modlong.bst}
\bibliography{refs}

\appendix

\section{Scheme dependence}
\label{app:scheme_dependence}
In this Appendix we quantify the dependence of the light vector meson production cross section on the choices for regularization and factorization schemes. The cross sections have been calculated as described in Sec.~\ref{sec:numerical_results}, and our default setup is the same. The dipole amplitude used in this Appendix is the KCBK evolved one from Ref.~\cite{Beuf:2020dxl} with the initial rapidity $\Ybk=4.61$.

\begin{figure}
        \centering
        \includegraphics[width=0.5\textwidth]{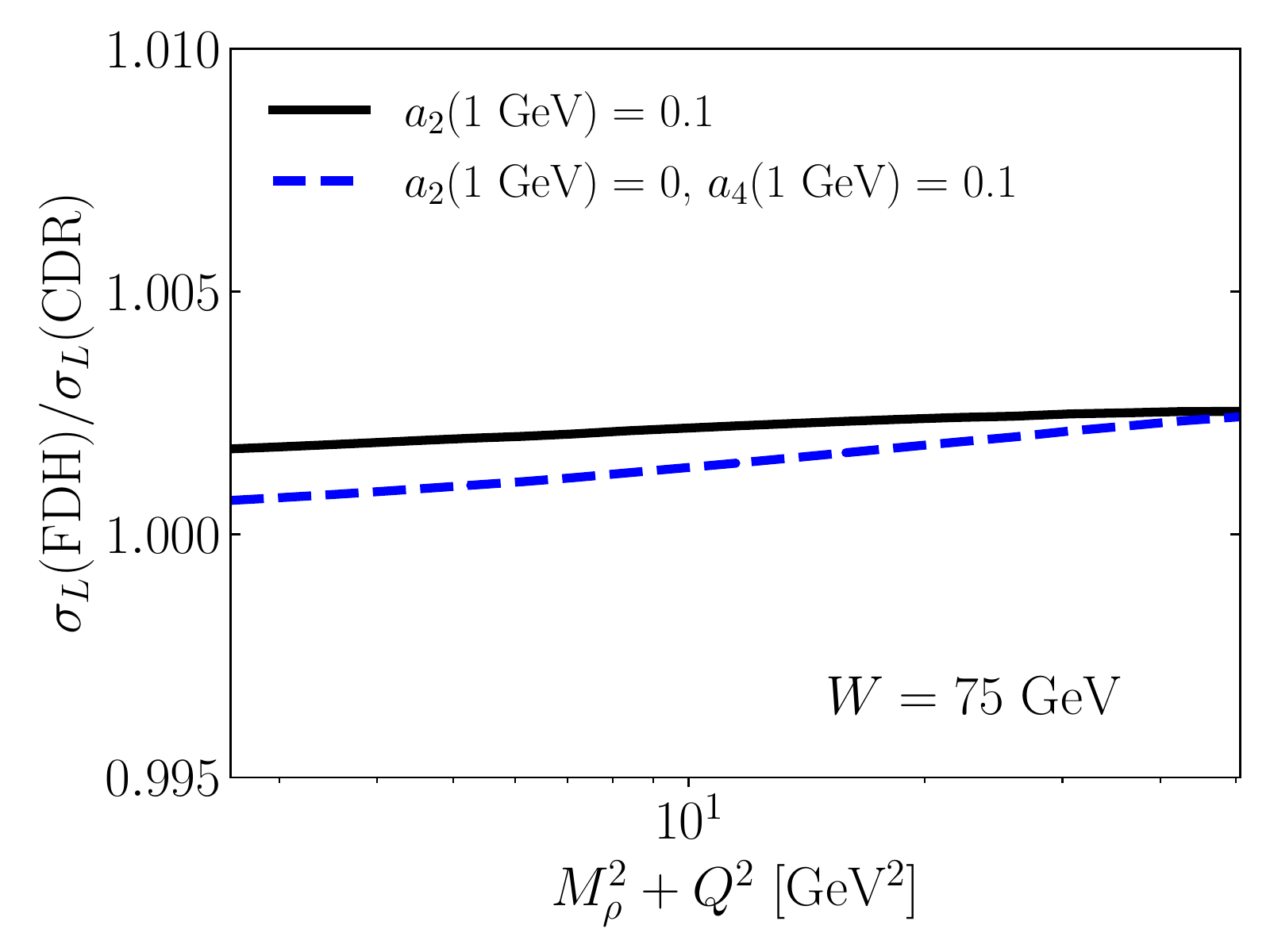}
        \caption{ Ratio of the $\rho$ production cross sections calculated in  FDH and CDR regularization schemes. 
        }
        \label{fig:reg_scheme_dependence}
\end{figure}

First, we show the dependence of the cross section on the %
regularization scheme in the transverse plane. The cross section has been calculated in both the CDR and FDH schemes and their ratio is shown in Fig.~\ref{fig:reg_scheme_dependence}. This ratio depends on the distribution amplitude, and it is equal to unity in the asymptotic limit $\mu_F^2 \rightarrow \infty$ where only the first term in the Gegenbauer expansion \eqref{eq:gegenbauer_expansion} contributes. For this reason we show the ratio with two different distribution amplitudes: one with $a_2(1\, \gev)= 0.1$ (our standard setup), and one with $a_2(1\, \gev)=0$, $a_4(1 \, \gev)=0.1$.  Higher-order terms are set to zero. We see that the dependence on the regularization scheme is very small in both cases, of the order $0.2 \%$ at most. 
Small scheme dependence is expected, as the first dominant term in the Gegenbauer expansion vanishes when one calculates the scheme dependent term in Eq.~\eqref{eq:NLO_qq_fin}.
It should be noted that the ratio does not seem to approach the asymptotic limit in the considered kinematics. This is a consequence of the ERBL evolution with running coupling being extremely slow, and the scheme dependence vanishes if we go to even higher values of virtuality. 

\begin{figure*}
	\centering
    \begin{minipage}{.5\textwidth}
        \centering
        \includegraphics[width=\textwidth]{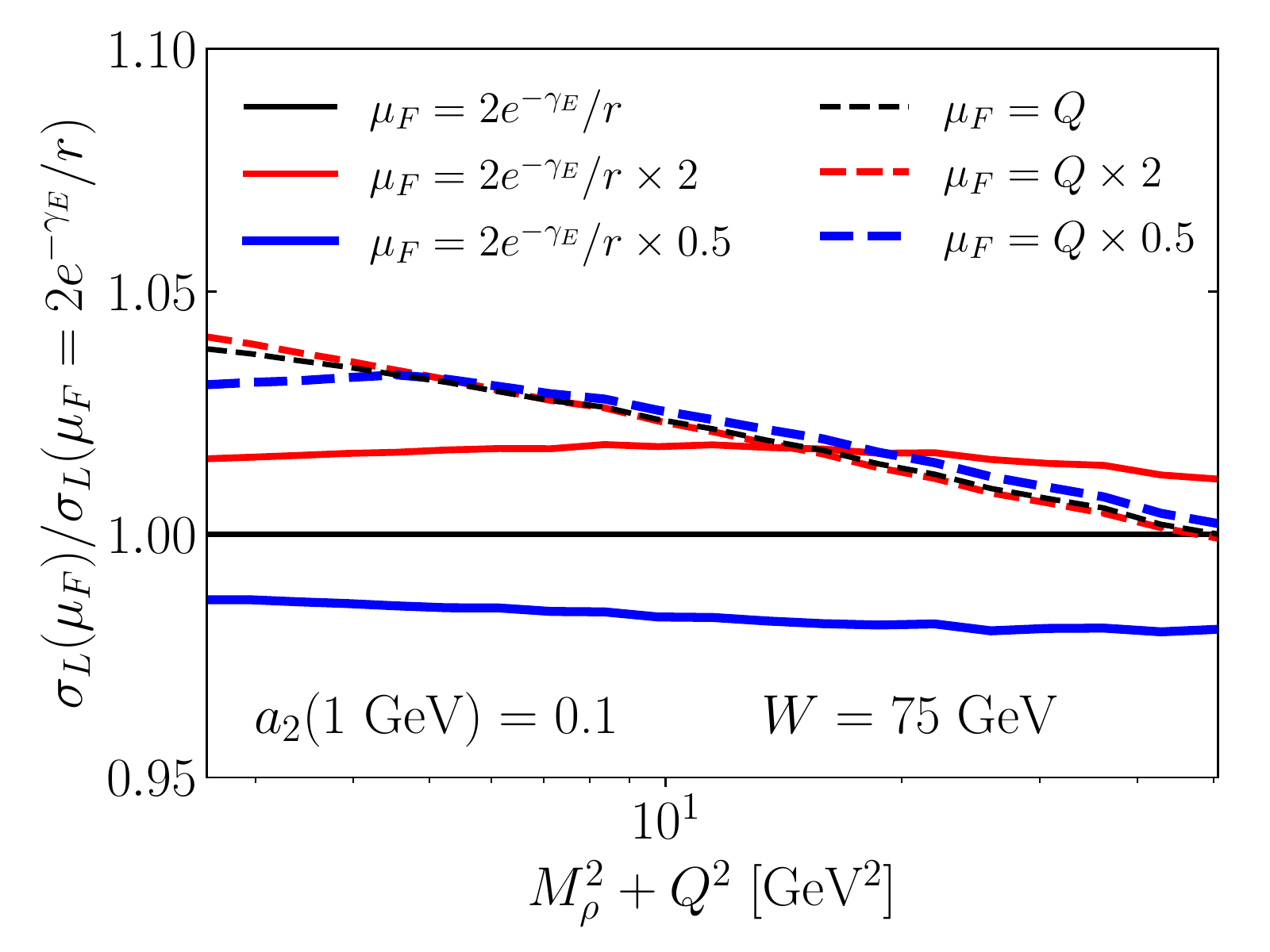}
        \caption{ Ratio of the cross section with different choices for the factorization scale $\mu_F$ to the default setup. }
        \label{fig:mu_F_scheme_dependence}
    \end{minipage}%
    \begin{minipage}{0.5\textwidth}
        \centering
        \includegraphics[width=\textwidth]{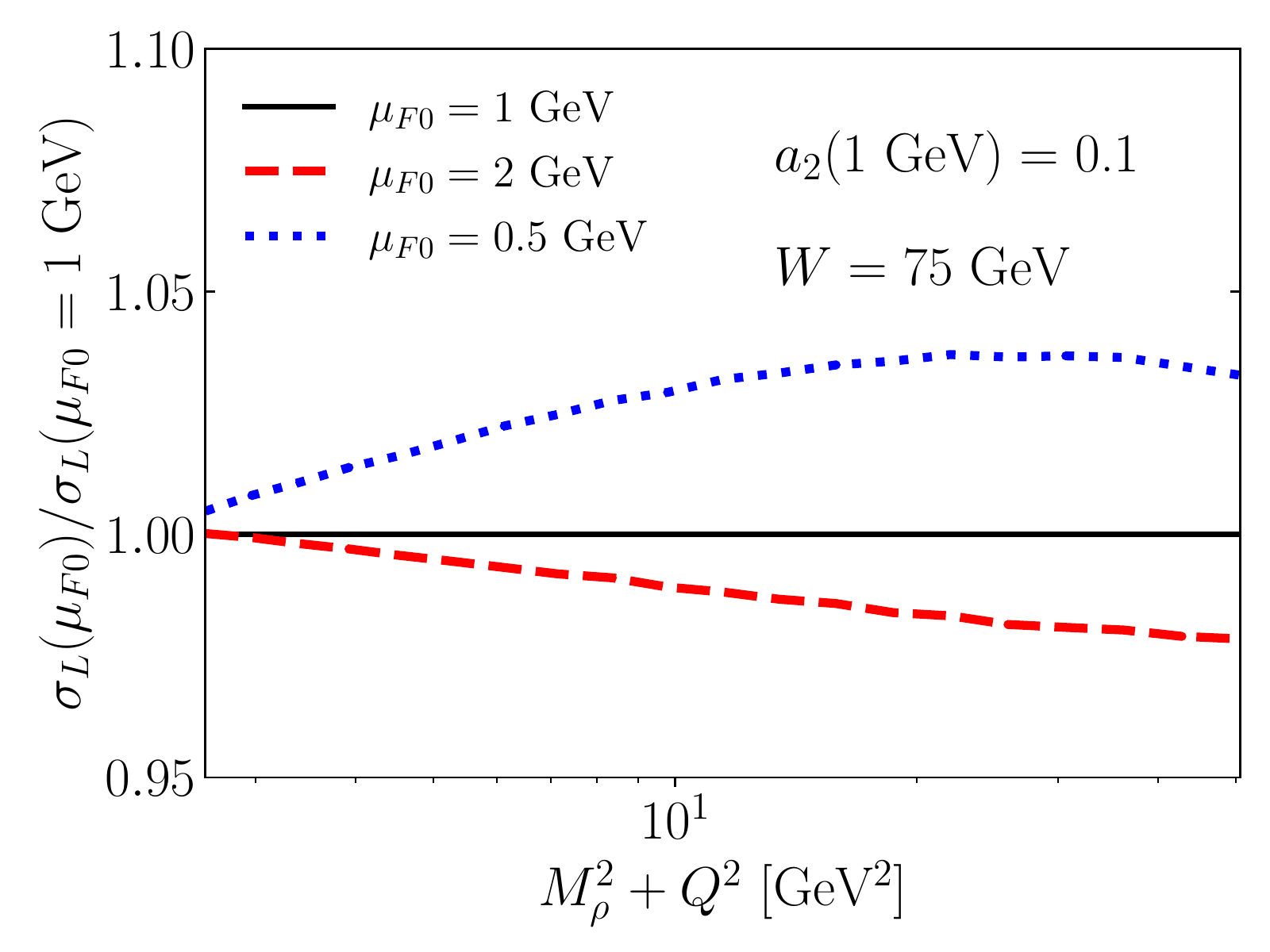}
        \caption{ Ratio of the cross section with different IR cutoffs $\mu_{F0}$ for the ERBL evolution to the default setup. }
        \label{fig:mu_F_cutoff_dependence}

    \end{minipage}
\end{figure*}

The cross section depends on the factorization scale $\mu_F$ at which the distribution amplitudes are evaluated as discussed in Sec.~\ref{sec:erbl}. In Fig.~\ref{fig:mu_F_scheme_dependence}, the cross sections have been calculated evaluating the distribution amplitude in both the $r$- and $Q$-schemes using our default choice for the distribution amplitude with $a_2(1\,\gev)=0.1$. The factorization scale has also been scaled by factors of $0.5$ and $2$. These results have been normalized by our default setup ($r$ scheme with $\mu_F=2e^{-\gamma_E}/r$). In the $Q$-scheme, varying the factorization scale by a factor $2$ has a very small effect. 
For the $r$-scheme the cross section varies somewhat more, but the variation is still only $\sim 2 \%$. The difference between the $r$- and $Q$-schemes is also only a few percent at most, meaning that the dependence on the factorization scale is small. This follows from the fact that the distribution amplitude receives only a small correction from the second scale-dependent Gegenbauer term, and the dominant term is factorization scale independent.

Finally, we show the dependence on the infrared cutoff $\mu_{F0}$ in Fig.~\ref{fig:mu_F_cutoff_dependence} using our default setup ($r$ scheme and $a_2(1\,\gev)=0.1$). There is some dependence on the IR cut-off, almost $5 \%$ at most with our choice for $a_2$. The reason for the cut-off dependence is that the dipole amplitude amplifies the contribution of large dipoles, meaning that dipoles of size $1/r \sim  1 \, \gev$ may have a numerically significant contribution even when $Q^2 \gg 1 \, \gev^2$. The dependence on the IR cut-off vanishes exactly
in the limit $Q^2 \rightarrow \infty$.

\end{document}